\documentclass[twocolumn,tighten]{aastex62}
\newcommand\iona[2]{#1$\;${\scshape{#2}}}
\newcommand{\mflux}{{erg~cm$^{-2}$~s$^{-1}$~Hz$^{-1}$}}
\newcommand{\flux}{{erg~cm$^{-2}$~s$^{-1}$}}
\newcommand{\lum}{{erg~s$^{-1}$}}
\newcommand{\lyanv}{Ly$\alpha$ + N~{\sc v}}

\begin{document}

\title{On the Fraction of \mbox{X-ray} Weak Quasars from the Sloan Digital Sky Survey}

\author{Xingting~Pu}
\affiliation{School of Astronomy and Space Science, Nanjing University,
Nanjing, Jiangsu 210093, China; bluo@nju.edu.cn}
\affiliation{College of Science, Nanjing Forestry University,
Nanjing, Jiangsu 210037, China}

\author{B.~Luo}
\affiliation{School of Astronomy and Space Science, Nanjing University,
Nanjing, Jiangsu 210093, China; bluo@nju.edu.cn}
\affiliation{Key Laboratory of Modern Astronomy and Astrophysics
(Nanjing University), Ministry of Education, Nanjing, Jiangsu 210093, China}
\affiliation{Collaborative Innovation Center of Modern Astronomy and Space Exploration,
Nanjing, Jiangsu 210093, China}

\author{W.~N.~Brandt}
\affiliation{Department of Astronomy \& Astrophysics, 525 Davey Lab,
The Pennsylvania State University, University Park, PA 16802, USA}
\affiliation{Institute for Gravitation and the Cosmos,
The Pennsylvania State University, University Park, PA 16802, USA}
\affiliation{Department of Physics, 104 Davey Lab,
The Pennsylvania State University, University Park, PA 16802, USA}

\author{John~D.~Timlin}
\affiliation{Department of Astronomy \& Astrophysics, 525 Davey Lab,
The Pennsylvania State University, University Park, PA 16802, USA}

\author{Hezhen~Liu}
\affiliation{School of Astronomy and Space Science, Nanjing University,
Nanjing, Jiangsu 210093, China; bluo@nju.edu.cn}
\affiliation{Key Laboratory of Modern Astronomy and Astrophysics
(Nanjing University), Ministry of Education, Nanjing, Jiangsu 210093, China}
\affiliation{Collaborative Innovation Center of Modern Astronomy and Space Exploration,
Nanjing, Jiangsu 210093, China}

\author{Q.~Ni}
\affiliation{Department of Astronomy \& Astrophysics, 525 Davey Lab,
The Pennsylvania State University, University Park, PA 16802, USA}
\affiliation{Institute for Gravitation and the Cosmos,
The Pennsylvania State University, University Park, PA 16802, USA}

\author{Jianfeng~Wu}
\affiliation{Department of Astronomy, Xiamen University, Xiamen, Fujian 361005, China}

\begin{abstract}
We investigate systematically the \mbox{X-ray} emission from type 1 quasars using a sample of 1825
Sloan Digital Sky Survey (SDSS) non-broad absorption line (non-BAL) quasars with {\it Chandra} archival observations.
A significant correlation is found between the \mbox{\mbox{X-ray}--to--optical} power-law slope parameter
($\alpha_{\rm OX}$) and the 2500~\AA\ monochromatic luminosity ($L_{\rm 2500~{\textup{\AA}}}$), and the \mbox{X-ray} weakness of
a quasar is assessed via the deviation of its $\alpha_{\rm OX}$ value from that expected from this relation.
We demonstrate the existence of a population of non-BAL \mbox{X-ray} weak quasars, and the fractions of quasars that are
\mbox{X-ray} weak by factors of $\ge6$ and $\ge10$ are \mbox{$5.8\pm0.7\%$} and \mbox{$2.7\pm0.5\%$},
respectively. We classify the \mbox{X-ray} weak quasars (\mbox{X-ray} weak by factors of $\ge6$) into three categories based on their optical
spectral features: weak emission-line quasars (WLQs; {\rm \iona{C}{iv}} REW $<16~{\textup{\AA}}$), red quasars ($\Delta(g-i)>0.2$),
and unclassified \mbox{X-ray} weak quasars. The \mbox{X-ray} weak fraction of $35_{- 9}^{+12}\%$ within the WLQ population
is significantly higher than that within non-WLQs, confirming previous findings that WLQs represent one population of
\mbox{X-ray} weak quasars. The \mbox{X-ray} weak fraction of $13_{- 3}^{+ 5}\%$ within the red quasar population
is also considerably higher than that within the normal quasar population. The unclassified \mbox{X-ray} weak quasars do
not have unusual optical spectral features, and their \mbox{X-ray} weakness may be mainly related to quasar \mbox{X-ray} variability.
\end{abstract}


\section{Introduction}

\mbox{X-ray} emission is an ubiquitous property of active galactic nuclei (AGNs). Extragalactic \mbox{X-ray} surveys with
{\it Chandra} and {\it XMM-Newton} have provided a quite complete understanding of the distant AGN population
(see \citealt{Brandt2015} for a review). AGN \mbox{X-ray} emission has been found to be
strongly correlated with the optical/UV emission \citep[e.g.,][]{AT1982,AT1986}.
It is believed that AGN optical/UV photons are emitted from the accretion disk, and
\mbox{X-ray} continuum emission arises from inverse Compton scattering of these optical/UV photons
in a hot accretion-disk ``corona'' \citep[e.g.,][]{Galeev1979,RN2003,Jiang2014}.
The \mbox{\mbox{X-ray}--to--optical} power-law slope parameter $\alpha_{\rm OX}$,
conventionally defined as \mbox{$\alpha_{\rm OX}=0.3838\log(L_{2~{\rm keV}}/L_{2500~{\textup{\AA}}})$}
\citep{Tananbaum1979},\footnote{$L_{2~{\rm keV}}$ and $L_{2500~{\textup{\AA}}}$
represent the \mbox{rest-frame} 2~keV and 2500~\AA\ monochromatic luminosities, respectively.}
is commonly used to quantify the ratio between the \mbox{X-ray} and optical/UV luminosity of AGNs.
Studies of the \mbox{X-ray} and optical/UV emission of large AGN samples have established a significant
\mbox{$\alpha_{\rm OX}$--$L_{\rm 2500~{\textup{\AA}}}$} anti-correlation spanning a broad range in
$L_{\rm 2500~{\textup{\AA}}}$ and redshift
\citep[e.g.,][]{Vignali2003,Strateva2005,Steffen2006,Just2007,Lusso2010,LR2016,Chiaraluce2018},
providing fundamental constraints on disk-corona models for AGNs.

The empirical \mbox{$\alpha_{\rm OX}$--$L_{\rm 2500~{\textup{\AA}}}$} relation
also allows identification of unusual \mbox{X-ray} weak AGNs (especially \mbox{X-ray} weak type 1 quasars), showing \mbox{X-ray}
emission significantly weaker than that expected from the \mbox{$\alpha_{\rm OX}$--$L_{\rm 2500~{\textup{\AA}}}$} relation.
The nearby ($z=0.192$) narrow-line type 1 quasar PHL~1811 is the best-studied example of this class \citep{Leighly2007b}.
The \mbox{X-ray} luminosity of PHL~1811 is a factor of $\sim$ 30--100 times weaker than that expected from
the \mbox{$\alpha_{\rm OX}$--$L_{\rm 2500~{\textup{\AA}}}$} relation. The steep \mbox{X-ray} spectrum
(with photon index $\Gamma=2.3\pm0.1$), lack of evidence for intrinsic \mbox{X-ray} and UV absorption,
and short-term \mbox{X-ray} flux variability by a factor of $\sim5$ strongly suggest that PHL~1811 is   
intrinsically \mbox{X-ray} weak \citep{Leighly2007b}. The optical/UV line emission of PHL~1811 is also unusual
(e.g., no forbidden or semiforbidden lines, very strong \iona{Fe}{ii} and \iona{Fe}{iii}, unusual
very low-ionization lines, and very weak high-ionization lines; see \citealt{Leighly2007a}).
Its \iona{C}{iv} emission line is blueshifted and asymmetric, and it has a
rest-frame equivalent width (REW) of only {6.6~\AA}, about 5 times smaller than
that in the composite quasar spectrum \citep{Vanden Berk2001}. 

Except for a few candidates for intrinsically \mbox{X-ray} weak quasars \citep[e.g.,][]{Gallagher2001,Wu2011,Luo2013,Luo2014,Liu2018}
like PHL~1811, \mbox{X-ray} weakness in \mbox{type 1} quasars is generally ascribed to absorption.
Broad absorption line (BAL) quasars are well known to be \mbox{X-ray} weak compared to quasars with
normal optical/UV spectra \citep[e.g.,][]{GM1996,Brandt2000,Gallagher2006}. Approximately one in every six
optically selected quasars shows BALs in its rest-frame UV spectra \citep[e.g.,][]{Reichard2003,Gibson2009a}.
Spectroscopic \mbox{X-ray} studies have found that BAL quasars typically have underlying \mbox{X-ray} continua
similar to those of normal quasars \citep{Gallagher2002} and the \mbox{X-ray} weakest BAL quasars tend to
have the hardest \mbox{X-ray} spectra \citep{Gallagher2006},
suggesting that the \mbox{X-ray} weakness in BAL quasars is primarily due to absorption.

There is a small population of weak emission-line quasars (WLQs; e.g.,
\citealt{Shemmer2009,DS2009,Plotkin2010a,LD2011,Wu2012}) that are known often to show weak \mbox{X-ray} emission.
They have broad UV emission lines (e.g., \iona{C}{iv}~$\lambda1549$) that are
significantly weaker than those of normal quasars. Studies of large samples of
this class of objects have found that a large fraction ($\ga$ 50\%) of WLQs are \mbox{X-ray} weak
\citep{Wu2012,Luo2015,Ni2018}. The small effective power-law photon index ($\Gamma_{\rm eff}\approx1.2$)
measured from \mbox{X-ray} stacking analyses indicates that the \mbox{X-ray} weak WLQs are on average likely
\mbox{X-ray} absorbed \citep[e.g., with $N_{\rm H}\approx9\times10^{23}$~{cm$^{-2}$};][]{Luo2015}. 

Few studies have investigated systematically the populations of \mbox{X-ray} weak quasars.
\citet{Gibson2008a} analyzed the \mbox{X-ray} and UV properties
of 536 Sloan Digital Sky Survey (SDSS; \citealt{York2000}) quasars,
including 315 with {\it Chandra} coverage and 221 with {\it XMM-Newton} coverage.
Based on 139 RQ non-BAL quasars in sample B of \citet{Gibson2008a}, their results showed that \mbox{X-ray} weak quasars are rare. 
Limited by their sample size, however, they only measured upper limits on the fraction of type 1 quasars
with a given factor of \mbox{X-ray} weakness (Figure 5 of \citealt{Gibson2008a}). 

Motivated by the significantly increased numbers of SDSS quasars and {\it Chandra} observations since
the \citet{Gibson2008a} work, we present here a systematic and uniform \mbox{X-ray} study of 1825 quasars with {\it Chandra} coverage,
which are drawn from the SDSS Seventh Data Release \citep[DR7;][]{Abazajian2009}
and Tenth Data Release \citep[DR10;][]{Ahn2014}.
Using the updated optical/UV and \mbox{X-ray} data, we can not only constrain better the fraction of \mbox{X-ray} weak quasars,
but also attempt to classify the causes for their \mbox{X-ray} weakness.
Studies of the fraction and nature of these populations of exceptional
objects may provide insights into the disk-corona system and nuclear obscuring material for
accreting supermassive black holes (SMBHs). We describe our sample selection and \mbox{X-ray}
data analysis in Sections \ref{sec:SA} and \ref{sec:X}, respectively. The results, including
the \mbox{X-ray} and optical/UV properties, and the fraction of \mbox{X-ray} weak quasars, are presented
in Section~\ref{sec:Re}. In Section~\ref{sec:Dis}, we discuss the nature of quasar \mbox{X-ray} weakness.
We summarize in Section~\ref{sec:SF}.
Throughout this work, we use J2000 coordinates and a cosmology with
$H_0=70$~km~s$^{-1}$~Mpc$^{-1}$, $\Omega_{\rm M}=0.3$,
and $\Omega_{\Lambda}=0.7$ \citep[e.g.,][]{Spergel2007}.

\section{Sample Selection} \label{sec:SA}

\subsection{SDSS and {\it Chandra} Archive Selection} \label{sec:AS}
The SDSS DR7 quasar catalog \citep{Schneider2010}
contains $105,783$ bona fide quasars in the redshift range
$0.065 < z < 5.460$ that have luminosities brighter than \mbox{$M_{i}=-22.0$} and
have at least one broad emission line with the full width at half-maximum (FWHM) larger than
1000~\mbox{km~s$^{-1}$} or have interesting/complex absorption features.
The SDSS DR10 quasar catalog \citep{Paris2014} contains $166,583$
Baryon Oscillation Spectroscopic Survey (BOSS) objects that
have luminosities \mbox{$M_{i}<-20.5$} and either have at least one emission line with the
FWHM larger than 500~\mbox{km~s$^{-1}$} or have interesting/complex absorption features.

We selected DR7 and DR10 quasars in the redshift range $1.700 < z < 2.700$.
The lower limit on the redshift ensures that the full velocity range up to 29,000~\mbox{km~s$^{-1}$} shortward of
the {\rm \iona{C}{iv}}~$\lambda1549$ emission line is redshifted into the SDSS and BOSS spectra, allowing unambiguous identification of
BAL quasars from their broad {\rm \iona{C}{iv}} absorption. The upper limit on the redshift ensures that the effective wavelengths
(rest-frame $\ga$ 2000~{\textup{\AA}}) of the absolute i-band magnitudes, which are extrapolated to derive the rest-frame 2500~\AA\ flux
densities (Section~\ref{sec:fo} below), do not deviate much from 2500~\AA.
We then searched the {\it Chandra} archive to find ACIS 
\citep[Advanced CCD Imaging Spectrometer;][]{Garmire2003} observations (with no gratings)
that were public as of 2016 July 1 and have pointing positions within
17\arcmin\ of the selected sources. We used the {\sc find\_chandra\_obsid} script in {\it Chandra} Interactive Analysis
of Observations (CIAO)\footnote{See \url{http://cxc.cfa.harvard.edu/ciao/} for details.}
to check if these sources are actually covered by {\it Chandra},\footnote{
If a source has multiple {\it Chandra} observations,
we selected the observation with the longest exposure and the smallest off-axis angle.
Sources that lie within 32 pixels of any chip edge were excluded.} 
resulting in a parent sample of 2475 SDSS quasars
with 1472 {\it Chandra} observation IDs.\footnote{There are 610
{\it Chandra} pointings that have two or more quasars within the field.}

\subsection{Excluding BAL Quasars} \label{sec:BAL}
It is necessary to exclude BAL quasars to avoid quasars with possible strong \mbox{X-ray} absorption.
The \citet{Shen2011} DR7 quasar catalog lists a {\tt BAL\_flag} parameter indicating the identification of BAL quasars.
This parameter flags both the objects in the SDSS Fifth Data Release (DR5) BAL quasar catalog \citep{Gibson2009a} and
the visually confirmed post-DR5 BAL quasars. The \citet{Paris2014} DR10 quasar catalog lists a {\tt BAL\_FLAG\_VI} parameter 
flagging those visually confirmed BAL quasars and also a {\tt BI\_CIV} parameter indicating the traditional BAL quasars
\citep{Weymann1991}. We consider a DR7 quasar with {\tt BAL flag} $>0$ in the \citet{Shen2011} catalog or a DR10 quasar with
{\tt BAL\_FLAG\_VI} $=1$ or {\tt BI\_CIV} $>0$ in the \citet{Paris2014} catalog a BAL quasar. In our parent sample,
345 quasars satisfy one or more of these criteria and they were thus excluded.

The $\rm BI_0$ definition adopted by \citet{Gibson2009a} and the BI definition adopted by \citet{Paris2014}
both apply a conservative absorption trough width threshold of 2000~\mbox{km~s$^{-1}$}. This threshold may miss relatively weak
BAL features. Therefore, we adopted the absorption index (AI; e.g., \citealt{Trump2006}) parameter to search for additional BAL
quasars. It was computed as
\begin{eqnarray}
{\rm AI} &\equiv& \int^{29,000}_{0}(1-f(v))C~dv,
\end{eqnarray}
where $f(v)$ is the continuum-normalized flux density. The value of $C$ is initially set to zero;
it is set to 1 whenever $f(v)$ has been continuously less than 0.9 for more than
1000~\mbox{km~s$^{-1}$}. A sample object with ${\rm AI}>0$ is also considered a BAL quasar.

We fitted the SDSS and BOSS spectra of our sample quasars following the method of
\citet{Gibson2008b}, which we summarize briefly here.
For each sample quasar, we corrected its spectrum for Galactic extinction using the reddening curve of
\citet{Cardelli1989} and \citet{Odonnell1994}. The continuum was fitted by a power law with Small Magellanic Cloud
(SMC) reddening \citep{Pei1992}; the continuum regions were selected to be rest-frame 1250--1350, 1600--1800,
1950--2050, 2150--2250, and 2950--3700~\AA, which are free from strong emission and absorption features.
We iteratively fitted the continuum regions with a 3$\sigma$ clipping method (i.e., spectral bins that deviate from
the continuum model by more than 3$\sigma$ were ignored at each iteration, with $\sigma$ being the spectral error).
The continuum was determined using the three best-fit parameters: power-law normalization,
spectral index $\alpha_{\lambda}$, and reddening $E(B-V)$. Since the spectral index $\alpha_{\lambda}$ and
reddening $E(B-V)$ are degenerate, the $E(B-V)$ value derived from the fitting is not necessarily
attached to any physical significance. Thus, the continuum is not corrected for any intrinsic reddening.
To identify \iona{C}{iv} BAL features, we smoothed the SDSS or BOSS spectrum using a 3 pixel boxcar and
computed the AI parameter. For a quasar that has both SDSS and BOSS spectra, we consider it a BAL quasar if either
of its spectra has ${\rm AI}>0$. Using this criterion, 173 additional BAL quasars were excluded,
leaving us with 1957 quasars in the parent sample.

\subsection{Measuring $f_{\rm 2500~{\textup{\AA}}}$} \label{sec:fo}
The measurement of the flux density at \mbox{rest-frame} 2500~\AA\ ($f_{\rm 2500~{\textup{\AA}}}$) is key for our analyses.
In this paper, we derived the $f_{\rm 2500~{\textup{\AA}}}$ values from the 2500~\AA\
luminosities, which were converted from the absolute {\it i}-band magnitudes $M_{i}(z=2)$ using Equation~4 of \citet{Richards2006}.
The $M_{i}(z=2)$ values were adopted from the \citet{Shen2011} DR7 and \citet{Paris2014} DR10 quasar catalogs. 
Since the absolute flux calibration errors in the BOSS spectra are relatively large and wavelength dependent
\citep[e.g.,][]{Paris2012,Margala2016}, we chose to estimate the $f_{\rm 2500~{\textup{\AA}}}$ values from $M_{i}(z=2)$ instead of
from spectral fitting for both SDSS DR7 and BOSS DR10 quasars to ensure homogeneous flux measurements.

We compared our $f_{\rm 2500~{\textup{\AA}}}$ values for the DR7 quasars in our final sample (see Section~\ref{sec:FS} below)
to those reported in the \citet{Shen2011} catalog. The \citet{Shen2011} $f_{\rm 2500~{\textup{\AA}},Shen}$ values,
which were measured from spectral fitting, are largely consistent with our $f_{\rm 2500~{\textup{\AA}}}$ values,
with a mean $f_{\rm 2500~{\textup{\AA}}}/f_{\rm 2500~{\textup{\AA}},Shen}$ value of 0.99 and a standard deviation of 0.35. In addition,
we verified that our $f_{\rm 2500~{\textup{\AA}}}$ values are consistent with those ($f_{\rm 2500~{\textup{\AA}},spec}$) derived from
our continuum fitting results in Section~\ref{sec:BAL} (i.e., using power-law normalization, spectral index $\alpha_{\lambda}$,
and reddening $E(B-V)$) for the DR7 quasars in our final sample, with a mean $f_{\rm 2500~{\textup{\AA}}}/f_{\rm 2500~{\textup{\AA}},spec}$
value of 0.98 and a standard deviation of 0.28. We not that since the photometry and spectra are taken at different times, some of the
standard deviation may just be due to quasar variability (rather than our methodological errors).
For the DR10 quasars in our final sample (having BOSS spectra),
we identified small wavelength dependent calibration offsets
between the photometric magnitudes and the spectrum synthesized magnitudes \citep[e.g.,][]{Margala2016},
indicating that any flux density measurements
from the continuum fitting results in Section~\ref{sec:BAL} may be unreliable.\footnote{We note that such small
calibration errors hardly affect the computations of the AI parameters in Section~\ref{sec:BAL} or the BI parameters in
\citet{Paris2014}, and thus we do not consider the BAL quasar exclusion unreliable.}

\subsection{Excluding RL Quasars} \label{sec:RL}
We also need to exclude radio-loud (RL) quasars, since they are well known to have \mbox{X-ray} emission levels systematically
higher than those of radio-quiet (RQ) quasars with comparable optical/UV luminosities \citep[e.g.,][]{Zamorani1981,Worrall1987}.
\mbox{X-ray} emission related to the jet, in addition to enhanced emission from the accretion-disk corona, may be required to interpret the
\mbox{X-ray} excess of RL quasars \citep{Miller2011,Zhu2020}.
Following \citet{Jiang2007}, we matched the remaining sample of 1957 quasars to the latest
Faint Images of the Radio Sky at Twenty-cm \citep[FIRST;][]{Beck1995} survey catalog (14Dec17 version)\footnote{
\url{http://sundog.stsci.edu/.}} with a matching radius of 30\arcsec. We classified objects that have only one radio component
within 5\arcsec\ as core-dominated quasars, and objects that have multiple radio components within 30\arcsec\ as
lobe-dominated quasars. The sample contains 99 FIRST-detected quasars, including 80
core-dominated quasars and 19 lobe-dominated quasars. For each FIRST-detected object, we used the integrated flux density listed
in the FIRST catalog to calculate its 1.4~GHz flux. For each lobe-dominated quasar, the total radio flux was calculated using
all of its radio components within 30\arcsec. We visually examined the quasar optical image from the Digital Sky Survey (DSS)
to ensure that its radio components are not associated with any interloper. We also matched the 50 quasars
that lie outside the FIRST survey area to the NRAO VLA Sky Survey (NVSS; \citealt{Condon1998}).
Three objects have one NVSS source matched within 30\arcsec. The upper limits on the 1.4~GHz fluxes for FIRST-undetected objects
were set to \mbox{$0.25+3\sigma_{\rm rms}$}~mJy, where 0.25 is the {\sc clean} bias correction and $\sigma_{\rm rms}$ is the
median rms noise of the FIRST survey \citep[0.14;][]{White1997}; the upper limits on the 1.4~GHz fluxes for NVSS-undetected objects
were set to 1.35~mJy, corresponding to three times the typical rms noise of the NVSS.

The radio-loudness parameter is defined as \linebreak \mbox{$R=f_{5~{\rm GHz}}/f_{\rm 2500~{\textup{\AA}}}$} \citep[e.g.,][]{Stocke1992},
where $f_{5~{\rm GHz}}$ is the flux density at rest-frame 5~GHz. The $f_{5~{\rm GHz}}$ values (or their upper limits)
were converted from the observed 1.4~GHz flux densities assuming a radio power-law slope of $\alpha_{\rm r}=-0.5$.
The FIRST survey is not sufficiently sensitive to discriminate $R \ge 10$ vs. $R<10$ for a large fraction of our sample objects,
and we therefore classify objects with $R \geq 100$ as RL, objects with $10\leq R <100$ as radio-intermediate (RI),
and objects with $R<10$ as RQ. There are 413 RQ, 30 RI, and 59 RL quasars within the remaining sample.
Another 1455 objects with upper limits on $R$ larger than 10 are referred to as radio-unclassified (RU),
97 of which have upper limits on $R$ larger than 100. We excluded the 59 RL quasars,
leaving 1898 objects in our sample.

\begin{figure}[htb!]
\centerline{
\includegraphics[scale=0.5]{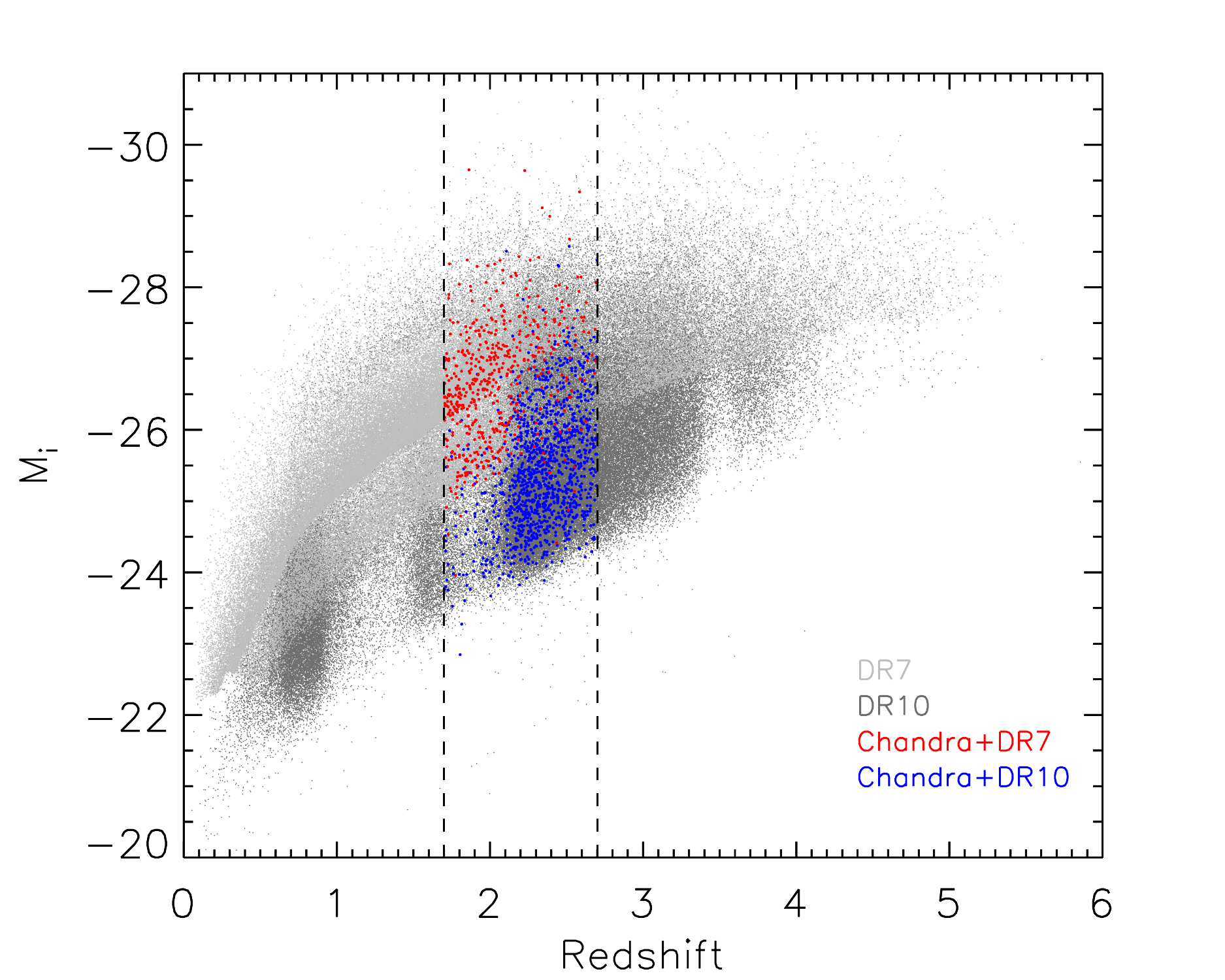}
}
\caption{Redshift vs.\ absolute {\it i}-band magnitude for the 1825 quasars in the final sample (sample A);
the red (blue) dots represent sample objects in the SDSS DR7 (DR10) catalog. The vertical dashed lines mark the redshift
boundaries of $z=1.700$ and $z=2.700$. The underlying light and dark gray dots represent all the SDSS DR7 and DR10 quasars,
respectively.
\label{fig:Miz} }
\end{figure}

\subsection{Final Sample} \label{sec:FS}
After excluding the BAL and RL quasars from the parent sample, we also removed 73 quasars that are {\it Chandra} targets;
these \mbox{X-ray} targets might have unusual \mbox{X-ray} properties and they may bias our systematic investigation of \mbox{X-ray}
weak quasars. The final sample contains 1825 quasars, and we refer to this sample as sample A. A summary of the
sample selection is presented in Table~1. Sample A contains 555 DR7 quasars and 1270 DR10 quasars,\footnote{
There are 137 quasars in sample A that are in both the DR7 and DR10 quasar catalogs. We refer to these quasars
as DR7 quasars, and we preferentially adopt their properties from the DR7 catalog.}
and we show in Figure~\ref{fig:Miz} the 1825 sample A objects together with all the SDSS DR7 and DR10 quasars
in the redshift versus absolute {\it i}-band magnitude $M_{i}(z=2)$ plane.

\begin{deluxetable}{lcc}
\tabletypesize{\scriptsize}
\tablecaption{Sample Selection}
\tablehead{
\colhead{Sample}                 &
\colhead{Selection Criteria}     &
\colhead{Number of Quasars}     \\
}
\startdata
& in DR7 or DR10 quasar catalog &  \\
Parent Sample & $1.700 < z < 2.700$ & 2475 \\
& in {\it Chandra} Archive ($07/01/2016$) &  \\
\hline
& in Parent Sample &  \\
Sample A & excluding 518 BAL quasars & 1825 \\
& excluding 59 RL quasars &  \\
& excluding 73 {\it Chandra} targets &  \\
\hline
& in Sample A &  \\
& in DR7 quasar catalog &  \\
Sample B & $m_{i}<19.6$  & 218 \\
& effective exposure time $>$2.5~ks & \\
& off-axis angle $<$9.8\arcmin & \\
\hline
& in Sample A &  \\
& in DR10 quasar catalog &  \\
Sample C & $m_{i}<21.1$ & 208 \\
& effective exposure time $>$6.3~ks & \\
& off-axis angle $<$8.2\arcmin & \\
\enddata
\label{tbl:SS}
\end{deluxetable}

The \mbox{X-ray} observations for the sample A quasars are listed in Table~\ref{tbl:X}.
The measured $f_{\rm 2500~\textup{\AA}}$ and $R$ values (or their upper limits; see Section~\ref{sec:RL}), along with the
absolute $i$-band magnitudes $M_{i}$, REWs of the \iona{C}{iv}~$\lambda1549$ emission, and relative $g-i$ colors
(i.e., $\Delta(g-i)$; e.g., \citealt{Richards2003}) derived from the SDSS DR7 and DR10 quasar catalogs for the sample A objects
are listed in Table~\ref{tbl:XO}. Properties from the SDSS DR7 quasar catalog are preferred when a quasar
has an entry in both catalogs. We listed the \iona{C}{iv} REW and $\Delta(g-i)$ values, since weak \iona{C}{iv}
emission lines and red $\Delta(g-i)$ colors are probably related to quasar \mbox{X-ray} weakness (see Sections~\ref{sec:WLQ}
and \ref{sec:red} for further discussion). There are 31 DR10 quasars in sample A that have \iona{C}{iv}~$\lambda1549$ REW
values of $-1$ in the catalog, indicating that a principal component analysis (PCA) failed to fit the emission line
\citep{Paris2014}. For each of these objects, we fitted a power-law local continuum between rest-frame 1450--1470
and \mbox{1650--1820~\AA}, and then measured the \iona{C}{iv} REW between 1500 and 1600~\AA\ (see Section 7.3 of \citealt{Paris2012}).
The \iona{C}{iv} REW values of these 31 DR10 quasars range from 0.9 to $\rm {479.1~\AA}$, with a median value of $\rm {33.8~\AA}$
and a mean value of $\rm {84.5~\AA}$.\footnote{Adopting the same method, our measurements of the \iona{C}{iv} REWs for other DR10 quasars
in sample A are generally consistent with the \iona{C}{iv} REWs reported in the \citet{Paris2014} DR10 quasar catalog.}

\begin{deluxetable*}{lccrrccccccc}
\tabletypesize{\scriptsize}
\tablecaption{{\it Chandra} Observations and \mbox{X-ray} Photometric Properties}
\tablehead{
\colhead{SDSS Name}                 &
\colhead{Data Release}                  &
\colhead{Redshift}                  &
\colhead{Observation}                  &
\colhead{Exposure}                   &
\colhead{Broad Band}                   &
\colhead{Soft Band}                   &
\colhead{Hard Band}                   &
\colhead{Band}                   &
\colhead{$\Gamma_{\rm eff}$}     &
\colhead{Sample B}  &
\colhead{Sample C}  \\
\colhead{(J2000)}   &
\colhead{}   &
\colhead{}   &
\colhead{ID}   &
\colhead{Time (ks)}   &
\colhead{(0.5--7~keV)}   &
\colhead{(0.5--2~keV)}   &
\colhead{(2--7~keV)}                   &
\colhead{Ratio}                   &
\colhead{}   &
\colhead{}   &
\colhead{}   \\
\colhead{(1)}         &
\colhead{(2)}         &
\colhead{(3)}         &
\colhead{(4)}         &
\colhead{(5)}         &
\colhead{(6)}         &
\colhead{(7)}         &
\colhead{(8)}         &
\colhead{(9)}         &
\colhead{(10)}        &
\colhead{(11)}        &
\colhead{(12)}        
}
\startdata
$000015.47+005246.8$  &  $   7$  &  $ 1.8571$  &  $ 11591$  &  $ 21.7$  &  $  61.5_{ -8.6}^{ +9.7}$  &  $  40.1_{ -6.6}^{ +7.7}$  &  $  21.4_{ -5.5}^{ +6.6}$  &  $ 0.53_{  -0.15}^{  +0.18}$  &  $ 1.6_{-0.3}^{+0.3}$  &  $ 0$  &  $ 0$  \\
$000018.18+050803.6$  &  $  10$  &  $ 2.2279$  &  $  7334$  &  $  2.7$  &  $<   8.3$  &  $<   8.7$  &  $<   5.9$  &  $...$  &  $ 1.8$  &  $ 0$  &  $ 0$  \\
$000029.98+004845.3$  &  $  10$  &  $ 2.4120$  &  $ 11591$  &  $ 21.2$  &  $  11.3_{ -5.2}^{ +6.3}$  &  $   6.8_{ -3.4}^{ +4.5}$  &  $<  18.6$  &  $ 0.79_{  -0.60}^{  +1.48}$  &  $ 1.1_{-1.0}^{+1.4}$  &  $ 0$  &  $ 0$  \\
$000106.87+023845.9$  &  $  10$  &  $ 1.7600$  &  $  4837$  &  $  5.0$  &  $   4.5_{ -2.2}^{ +3.4}$  &  $   4.7_{ -2.2}^{ +3.4}$  &  $<   5.9$  &  $ 0.15_{  -0.14}^{  +0.22}$  &  $ 2.6_{-0.9}^{+2.0}$  &  $ 0$  &  $ 0$  \\
$000159.88+003715.5$  &  $  10$  &  $ 2.3910$  &  $  4861$  &  $  4.8$  &  $<  16.7$  &  $<  13.2$  &  $<  11.3$  &  $...$  &  $ 1.8$  &  $ 0$  &  $ 0$  \\
\hline
\enddata
\tablecomments{
Cols. (1)--(3): SDSS name in the J2000 equatorial coordinate format, SDSS Data Release number, and redshift.
The improved redshifts from \citet{HW2010} reported in the \citet{Shen2011} catalog are listed for the DR7 quasars.
Col. (4): {\it Chandra} observation ID. 
Col. (5): {\it Chandra} effective exposure time in the broad (0.5--7~keV) band.
Cols. (6)--(8): broad-band (0.5--7~keV), soft-band (0.5--2~keV), and hard-band (2--7~keV) net counts in the source region.
A 3$\sigma$ confidence level upper limit on the source counts is given if the source is undetected.
Col. (9): band ratio between the hard-band and soft-band counts. An entry
of ``...'' indicates that the source is undetected in both bands.
Col. (10): 0.5--7~keV effective power-law photon index derived from the band ratio.
It is fixed at 1.8 for any source that is undetected in both the soft and hard bands (see Section~\ref{sec:X}). 
Cols. (11) and (12): an entry of ``1'' indicates that the object is in sample B or sample C (see Section~\ref{sec:frac_BC}).
Table~\ref{tbl:X} is sorted by increasing J2000 right ascension and is published in its entirety in the electronic edition of the journal. 
A portion is shown here for guidance regarding its form and content.
}
\label{tbl:X}
\end{deluxetable*}

\begin{deluxetable*}{lcrrrcrrrccrc}
\tabletypesize{\scriptsize}
\tablecaption{\mbox{X-ray} and Optical Properties}
\tablehead{
\colhead{SDSS Name}                 &
\colhead{$M_{i}$}                  &
\colhead{$f_{\rm 2~keV}$}                  &
\colhead{$F_{\rm 0.5-7~keV}$}                  &
\colhead{$\log L_{\rm 2-10~keV}$}                  &
\colhead{$f_{\rm 2500~\textup{\AA}}$}                  &
\colhead{$\alpha_{\rm OX}$}                  &
\colhead{$\Delta\alpha_{\rm OX}$}                   &
\colhead{$f_{\rm weak}$}                  &
\colhead{REW \iona{C}{iv}} &
\colhead{$\Delta(g-i)$}  &
\colhead{$R$}  &
\colhead{\mbox{X-ray} Weak} \\
\colhead{(1)}         &
\colhead{(2)}         &
\colhead{(3)}         &
\colhead{(4)}         &
\colhead{(5)}         &
\colhead{(6)}         &
\colhead{(7)}         &
\colhead{(8)}         &
\colhead{(9)}         &
\colhead{(10)}         &
\colhead{(11)}         &
\colhead{(12)}         &
\colhead{(13)}         
}
\startdata
$000015.47+005246.8$ & $-26.19$ & $    6.36$ & $    4.26$ & $  44.75$ & $  0.52$ & $  -1.50$ & $   0.00$ & $     1.0$ & $  44.1$ & $  0.131$ & $<  12.0$ & $  0$  \\
$000018.18+050803.6$ & $-25.75$ & $<  10.85$ & $<   5.40$ & $< 45.06$ & $  0.25$ & $< -1.29$ & $<  0.17$ & $>    0.4$ & $  57.0$ & $  0.056$ & $<  26.5$ & $  0$  \\
$000029.98+004845.3$ & $-24.43$ & $    0.74$ & $    0.91$ & $  44.22$ & $  0.06$ & $  -1.51$ & $  -0.17$ & $     2.7$ & $  78.9$ & $  0.047$ & $< 106.0$ & $  0$  \\
$000106.87+023845.9$ & $-23.98$ & $    2.92$ & $    0.82$ & $  44.03$ & $  0.07$ & $  -1.31$ & $  -0.00$ & $     1.0$ & $  51.6$ & $ -0.085$ & $<  82.5$ & $  0$  \\
$000159.88+003715.5$ & $-24.88$ & $<   9.55$ & $<   4.57$ & $< 45.06$ & $  0.10$ & $< -1.16$ & $<  0.23$ & $>    0.3$ & $  43.4$ & $ -0.031$ & $<  68.9$ & $  0$  \\
\hline
\enddata
\tablecomments{
Cols. (1) and (2): SDSS name in the J2000 equatorial coordinate format and absolute $i$-band magnitude.
Col. (3): rest-frame 2~keV flux density in units of $10^{-32}$~\mflux.
A 3$\sigma$ confidence level upper limit on $f_{\rm 2~keV}$ is given if the source is undetected.
Cols. (4) and (5): observed-frame 0.5--7~keV flux in units of $10^{-14}$~\flux and logarithm of the rest-frame 2--10~keV luminosity in units of \lum,
both calculated using $f_{\rm 2~keV}$ and $\Gamma_{\rm eff}$ (see Table~\ref{tbl:X}). 
Col. (6): rest-frame 2500~\AA\ flux density in units of $10^{-27}$~\mflux.
Col. (7): measured \mbox{X-ray}--to--optical power-law slope $\alpha_{\rm OX}$.
A 3$\sigma$ confidence level upper limit on $\alpha_{\rm OX}$ is given if the source is undetected.
Col. (8): difference between the measured $\alpha_{\rm OX}$ and that expected
from our best-fit \mbox{$\alpha_{\rm OX}$--$L_{\rm 2500~{\textup{\AA}}}$} relation.
Col. (9): \mbox{X-ray} weakness factor measured from $\Delta\alpha_{\rm OX}$.
Col. (10): REW (in units of \AA) of the \iona{C}{iv}~$\lambda1549$ emission line.
Col. (11): relative $g-i$ color.
Col. (12): radio-loudness parameter.
Col. (13): an entry of ``1'' indicates that the object is \mbox{X-ray} weak. We adopted $\Delta\alpha_{\rm OX} = -0.3$
to be the threshold separating \mbox{X-ray} normal and \mbox{X-ray} weak quasars (see Section~\ref{sec:frac}).
Table~\ref{tbl:XO} is sorted by increasing J2000 right ascension and is published in its entirety in the electronic edition of the journal.
A portion is shown here for guidance regarding its form and content.
}
\label{tbl:XO}
\end{deluxetable*}

\section{\mbox{X-ray} Data Analysis} \label{sec:X}

The \mbox{X-ray} data reduction was performed using CIAO version 4.8 and CALDB version 4.7.0.
For each observation ID, we reprocessed the {\it Chandra} dataset
using the {\sc chandra\_repro} script.
Background flares were removed using the {\sc deflare}
script with an iterative 3$\sigma$ clipping method.

We made for each of the 1825 sample objects \mbox{X-ray} images in the 0.5--7~keV (broad), 0.5--2~keV (soft),
and 2--7~keV (hard) bands from the cleaned level~2 event file.
\mbox{X-ray} source detection was performed on the images using the {\sc wavdetect} \citep{Freeman2002} script with wavelet scales
of 1, 1.414, 2, 2.828, 4, 5.656, and 8 pixels, and a significance threshold of 10$^{-6}$. If a sample object is detected
by {\sc wavdetect} within 3.0\arcsec\ of the SDSS position, we used the {\sc wavdetect}
position as its \mbox{X-ray} position; otherwise, we used the SDSS
astrometry as the \mbox{X-ray} position. Aperture photometry was performed to extract source counts
in the broad, soft, and hard bands. We used circular and annular regions centered at the \mbox{X-ray} position
to extract source and background counts, respectively. The radius of the circle was chosen to
enclose 90\% of the point spread function (PSF) at 1 keV, while the inner and outer radii of the annulus
were chosen to be the source radius plus 15 and 50 pixels, respectively.   
We visually inspected the background-extraction region for each sample object.
In cases where any {\sc wavdetect} source contaminates the background region, we excluded its
elliptical {\sc wavdetect} region that contains the majority of the source counts from the annulus.
We manually changed the annulus to a pie--shaped region\footnote{See \url{https://cxc.cfa.harvard.edu/ciao/ahelp/dmregions.html}
for a description of the pie--shaped region.} whenever the background region is partially off-chip. 

\begin{figure*}[htb!]
\centerline{
\includegraphics[scale=0.5]{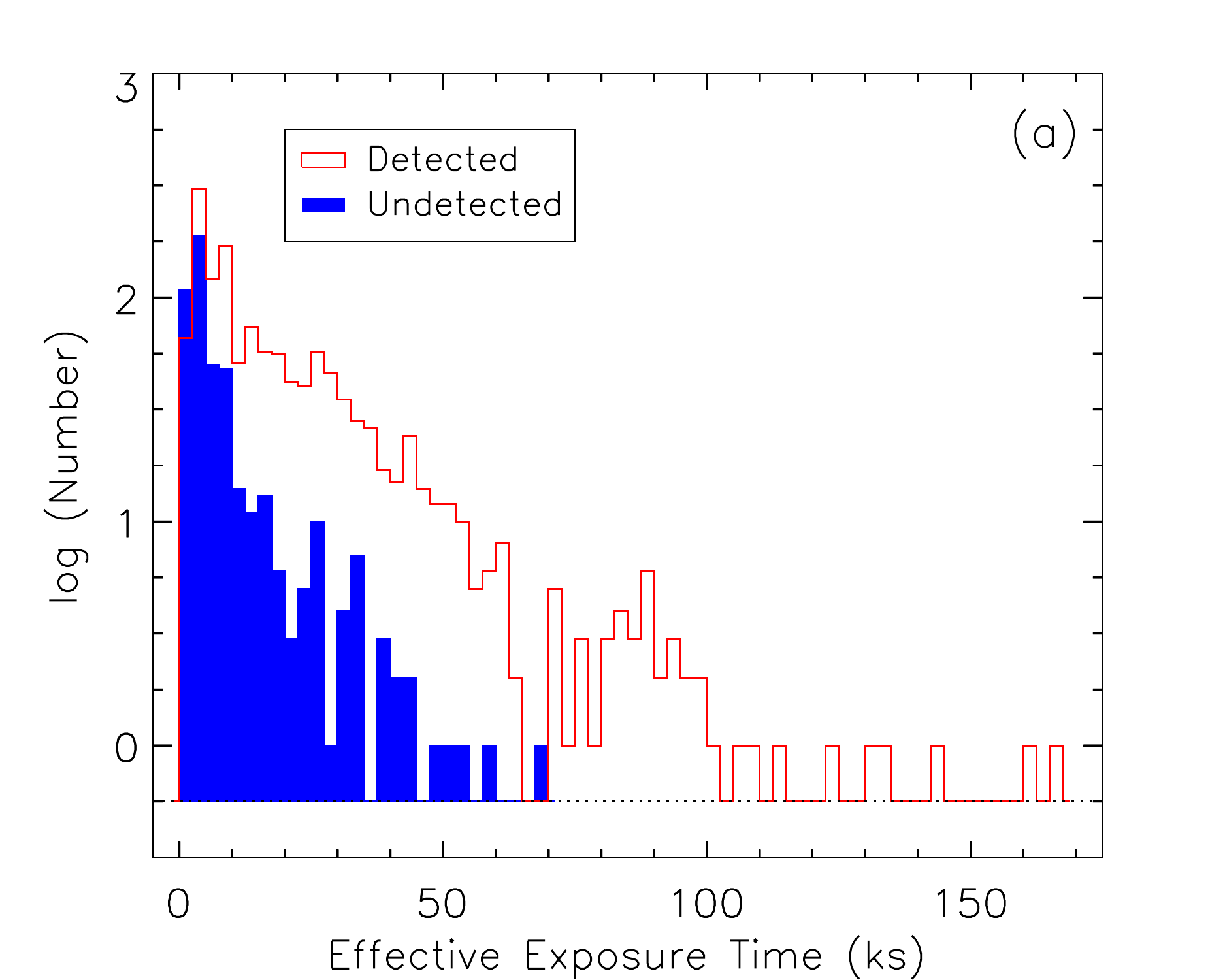}
\includegraphics[scale=0.5]{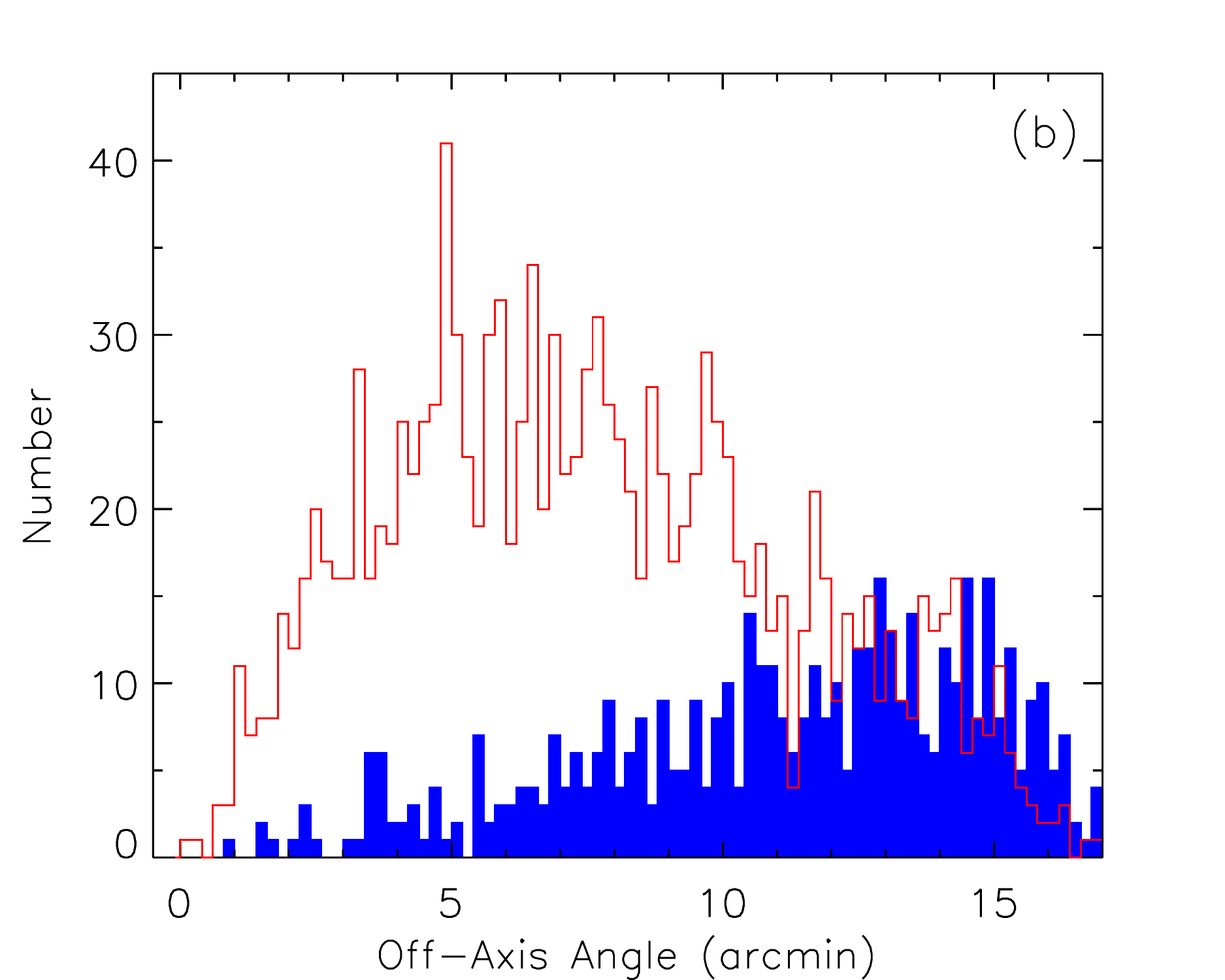}
}
\caption{Distributions of the (a) effective exposure time in the broad band (0.5--7~keV)
and (b) off-axis angle for sample A. The open red and solid blue histograms represent the \mbox{X-ray} detected
and \mbox{X-ray} undetected objects, respectively. In the left panel, the \mbox{{\it y}-axis} is logarithmic,
bins with zero counts are set to $-0.25$ (dotted horizontal line). In the right panel,
there is one sample object that has a small off-axis angle of $8.4\arcsec$. We note that 
it is not the target of the corresponding {\it Chandra} observation.
\label{fig:ta} }
\end{figure*}

To assess the detection significance, we calculated in each of the broad, soft, and hard bands a
binomial no-source probability $P_{\rm B}$ using the following equation \citep[e.g.,][]{Xue2011,Luo2013,Luo2015}:
\begin{equation}
P_{\rm B}(X\ge S)=\sum_{X=S}^{N}\frac{N!}{X!(N-X)!}p^X(1-p)^{N-X}~,
\end{equation}
where $S$ is the total counts in the source-extraction region; $N=S+B$, where $B$
is the total counts in the background-extraction region; $p\!=\!1/(1\!+\!BACKSCAL)$,
where $BACKSCAL$ is the background--to--source area ratio. If a sample object satisfied $P_{\rm B}< 0.01$ in one band,
we consider it detected in this band and calculated net counts along with associated 1$\sigma$ errors,
derived from 1$\sigma$ errors \citep{Gehrels1986} on the source and background counts;
otherwise we consider it undetected in this band and calculated a 3$\sigma$ confidence level upper limit on the source counts
using the Bayesian method of \citet{Kraft1991}.
We consider a source \mbox{X-ray} detected if it is detected in one or more of the three \mbox{X-ray} bands.
Our sample A contains 1344 \mbox{X-ray} detected and 481 \mbox{X-ray} undetected quasars.
Figure~\ref{fig:ta} shows the distributions of the broad-band effective exposure times and off-axis angles
for these \mbox{X-ray} detected and \mbox{X-ray} undetected objects. The effective exposure time is the
exposure time that has been corrected for the effects of vignetting and CCD gaps.
The \mbox{X-ray} detected quasars in general have longer exposures and smaller off-axis angles than the \mbox{X-ray}
undetected quasars. The numbers of detected/undetected DR7/DR10 quasars in specific \mbox{X-ray} bands are listed
in Table~\ref{tbl:NX}.

Assuming power-law \mbox{X-ray} spectra modified by Galactic absorption, we derived the effective power-law photon indices
($\Gamma_{\rm eff}$) from the band ratios for the 1285 sample objects that are detected in at least one of
the soft and hard bands. The band ratio is defined as the ratio between the hard-band and soft-band
counts. For each of the 732 objects that are detected in both the soft and hard bands, the band ratio was calculated
by dividing the hard-band net counts by the soft-band net counts, and its associated 1$\sigma$ uncertainties were
derived using the method of Bayesian estimation of hardness ratios \citep[BEHR;][]{Park2006}.
For each of the 553 objects that are detected in only one of the soft and hard bands,
we adopted the band ratio to be the mode of its posterior probability distribution computed using BEHR \citep{Park2006}.
These are best-guess estimates, and they appear more appropriate for deriving source fluxes than assuming a fixed spectral shape
\citep{Luo2017}. To derive $\Gamma_{\rm eff}$ from the band ratio,
we used the CIAO \mbox{\sc specextract} script to create the source spectrum, auxiliary response file (ARF), and
redistribution matrix file (RMF). We then used the Sherpa \citep[CIAO's modeling and fitting package; e.g.,][]{Freeman2001}
{\sc fake\_pha} command to create simulated spectra for a power-law plus Galactic absorption model with varying $\Gamma$. 
The observed band ratio was compared to the modeled band ratios, and the value of $\Gamma_{\rm eff}$
was determined by interpolating the modeled $\Gamma$ values.

For each of the 540 quasars that are undetected in both the soft and hard bands
(including 59 objects detected only in the broad band), the band ratio is not available;
thus we adopted a fixed $\Gamma_{\rm eff}$ value of 1.8, which is around the median value of $\Gamma_{\rm eff}$ for sample A objects
(see Section~\ref{sec:al} below). Among the 540 quasars with fixed $\Gamma_{\rm eff}$ values, there are three
\mbox{X-ray} detected (i.e., detected only in the broad band) objects having $\Delta\alpha_{\rm OX}\leq-0.3$ and
one \mbox{X-ray} undetected object having an upper limit on
$\Delta\alpha_{\rm OX}$ $\leq-0.3$; these four objects are considered \mbox{X-ray} weak (see Section~\ref{sec:frac} below).
Since the detected \mbox{X-ray} weak quasars have a median $\Gamma_{\rm eff}$ value of 0.8 (see Section~\ref{sec:frac} below),
these four quasars likely have smaller $\Gamma_{\rm eff}$ values.
However, as their number is small and they may also be intrinsically \mbox{X-ray} weak quasar candidates, we still chose to adopt
$\Gamma_{\rm eff}=1.8$ to calculate the relevant parameters (or their upper limits); adopting smaller $\Gamma_{\rm eff}$ values
(e.g., 0.8) for these four quasars would not affect significantly our results. The \mbox{X-ray} properties for sample A are listed
in Table~\ref{tbl:X}.

\begin{deluxetable*}{l|ccc|c|c}
\tabletypesize{\normalsize}
\tablecaption{Numbers of \mbox{X-ray} Detected/Undetected DR7/DR10 Quasars in Samples A, B, and C}
\tablehead{
\multicolumn{1}{c|}{\mbox{X-ray} Detection Breakdown}    &
\multicolumn{3}{c|}{Sample A} &
\multicolumn{1}{c|}{Sample B}   &
\colhead{Sample C}    \\
\multicolumn{1}{c|}{}    &
\colhead{DR7}    &
\colhead{DR10}    &
\multicolumn{1}{c|}{All}   &
\multicolumn{1}{c|}{DR7}    &
\colhead{DR10}  
}
\startdata
Detected in the Broad Band & $         473$ & $         806$ & $        1279$ & $         216$ & $         207$ \\
Detected in the Soft Band & $         475$ & $         781$ & $        1256$ & $         214$ & $         198$ \\
Detected in the Hard Band & $         321$ & $         440$ & $         761$ & $         179$ & $         158$ \\
Detected in \textit{Both} the Soft and Hard Bands & $         317$ & $         415$ & $         732$ & $         177$ & $         155$ \\
Detected in \textit{Only One} of the Soft and Hard Bands & $         162$ & $         391$ & $         553$ & $          39$ & $          46$ \\
\mbox{X-ray} Detected & $         490$ & $         854$ & $        1344$ & $         218$ & $         208$ \\
\mbox{X-ray} Undetected & $          65$ & $         416$ & $         481$ & $           0$ & $           0$ \\
\enddata
\tablecomments{We consider a source \mbox{X-ray} detected if it is detected in at least one of the three \mbox{X-ray} bands and consider
a source \mbox{X-ray} undetected if it is undetected in all three \mbox{X-ray} bands. All objects in samples B and C are \mbox{X-ray} detected.}
\label{tbl:NX}
\end{deluxetable*}

\section{Results} \label{sec:Re}

\subsection{\mbox{$\alpha_{\rm OX}$--$L_{\rm 2500~{\textup{\AA}}}$} Relation} \label{sec:al}

The \mbox{X-ray}--to--optical power-law slope parameter $\alpha_{\rm OX}$ or its upper limit was measured for each of our
sample objects. For each of the 1344 \mbox{X-ray} detected objects,
we derived its rest-frame 2~keV flux density ($f_{\rm 2~keV}$) by normalizing the broad-band counts
in the simulated spectrum with the derived $\Gamma_{\rm eff}$ value (see Section~\ref{sec:X}) to the
observed broad-band net counts. If a sample object is undetected in the broad band but detected in the soft (or hard) band,
we normalized the soft-band (or hard-band) counts in the simulated spectrum to the observed net counts. 
For each of the 481 \mbox{X-ray} undetected objects,
a 3$\sigma$ confidence level upper limit on $f_{\rm 2~keV}$ was computed by normalizing the broad-band counts in the simulated spectrum
to the upper limit on the observed broad-band counts (Section~\ref{sec:X}). 
The 3$\sigma$ upper limit is appropriate for the following survival analysis.
The $f_{\rm 2~keV}$ and $\alpha_{\rm OX}$ parameters are listed in Table~\ref{tbl:XO}.

\begin{figure}[htb!]
\centerline{
\includegraphics[scale=0.5]{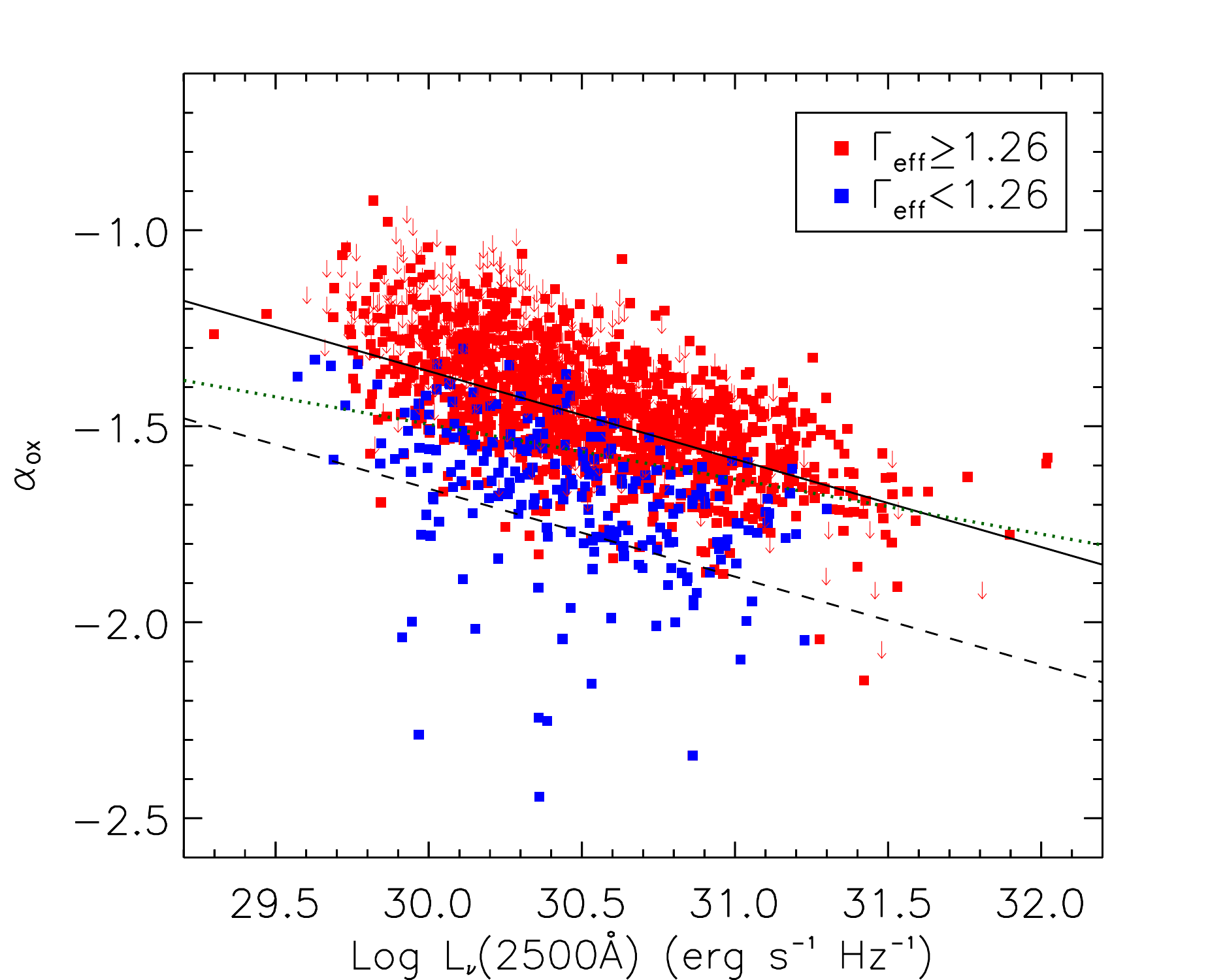}
}
\caption{$\alpha_{\rm OX}$ vs.\ 2500~\AA\ monochromatic luminosity for sample A. 
The red and blue filled squares represent \mbox{X-ray} unabsorbed ($\Gamma\ge1.26$)
and potentially \mbox{X-ray} absorbed objects ($\Gamma<1.26$), respectively;
downward arrows represent the $3\sigma$ upper limits on $\alpha_{\rm OX}$.
The solid black line shows the best-fit relation for the 1595 $\Gamma\ge1.26$ objects,
while the dashed black line shows $\Delta\alpha_{\rm OX}=-0.3$ from the best-fit relation,
separating \mbox{X-ray} normal and \mbox{X-ray} weak quasars (see Section~\ref{sec:frac} below).
The green dotted line shows the \citet{Just2007} \mbox{$\alpha_{\rm OX}$--$L_{\rm 2500~{\textup{\AA}}}$} relation
for comparison.
\label{fig:al_0a} }
\end{figure}

Previous studies have revealed a strong anticorrelation between $\alpha_{\rm OX}$ and 2500~\AA\ monochromatic luminosity 
over five orders of magnitude in $L_{\rm 2500~{\textup{\AA}}}$ \citep[e.g.,][]{Wilkes1994,Strateva2005,Steffen2006,Just2007}.
Figure~\ref{fig:al_0a} shows the $\alpha_{\rm OX}$ versus $L_{\rm 2500~{\textup{\AA}}}$ distribution for our sample A quasars. 
The generalized Kendall's $\tau$ test in the Astronomy Survival Analysis package \citep[ASURV Rev 1.2;][]{IF1990,LaValley1992}
confirmed a highly significant (20.4$\sigma$) anticorrelation between $\alpha_{\rm OX}$ and $L_{\rm 2500~{\textup{\AA}}}$.

\begin{figure*}[htb!]
\centerline{
\includegraphics[scale=0.5]{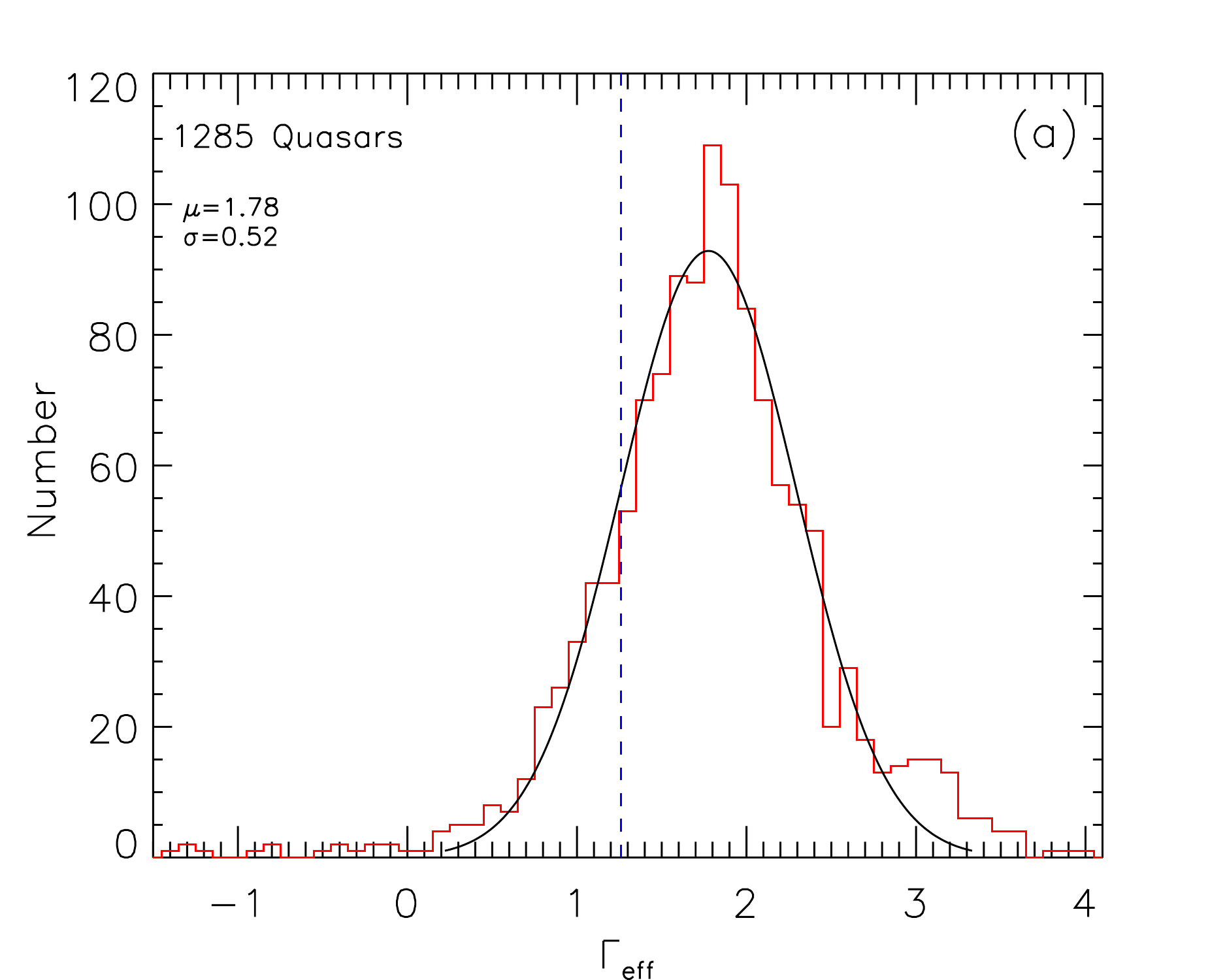}
\includegraphics[scale=0.5]{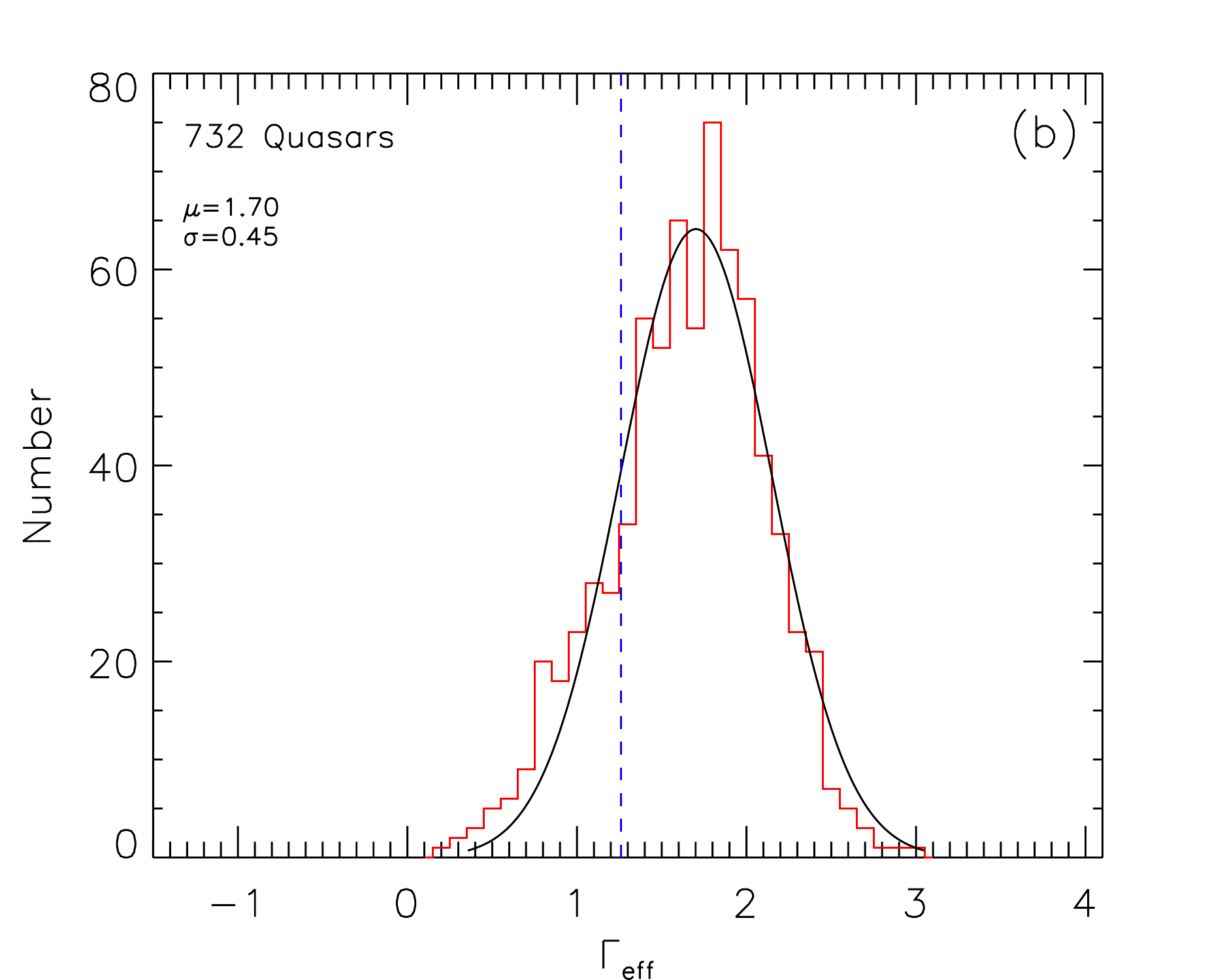}
}
\caption{Distribution of $\Gamma_{\rm eff}$ for (a) 1285 quasars that are detected in at least one of the soft and hard bands,
and (b) 732 quasars that are detected in both the soft and hard bands. In each panel,
the black curve shows the best-fit Gaussian profile, with the mean and standard deviation of the Gaussian fit listed
in the upper left corner; the vertical dashed line indicates $\Gamma_{\rm eff}=1.26$,
which is the threshold adopted in this study to separate \mbox{X-ray} unabsorbed and potentially \mbox{X-ray} absorbed quasars.
\label{fig:geff} }
\end{figure*}

Before quantifying the \mbox{$\alpha_{\rm OX}$--$L_{\rm 2500~{\textup{\AA}}}$} relation, we attempted to exclude
potentially \mbox{X-ray} absorbed objects. We selected potentially \mbox{X-ray} absorbed objects based on
the effective power-law photon index $\Gamma_{\rm eff}$. The mean value of $\Gamma_{\rm eff}$ for RQ type 1 quasars
has been found to be $\Gamma\approx1.9$ \citep[e.g.,][]{Reeves1997,Just2007}. The $\Gamma_{\rm eff}$ value generally becomes smaller
if a quasar is \mbox{X-ray} absorbed. For the 1285 quasars in our sample that are detected in at least one of the soft and
hard bands, the distribution of $\Gamma_{\rm eff}$ is shown Figure~\ref{fig:geff}(a). 
The $\Gamma_{\rm eff}$ values range from $-1.4$ to 4.0 with an average value of 1.8 and a standard deviation of 0.7,
consistent with those derived in previous studies \citep[e.g.,][]{Just2007}.

We fitted the $\Gamma_{\rm eff}$ distribution with a Gaussian function using the IDL MPFIT \citep{2009ASPC..411..251M} routine.
The black curve in Figure~\ref{fig:geff}(a) is the best-fit Gaussian profile, which has a mean of 1.78 and a standard deviation
of 0.52. We note that there are 553 objects among these 1285 quasars that are detected in only one of the soft and hard bands,
and their $\Gamma_{\rm eff}$ values are best-guess estimates (Section~\ref{sec:X}).
The $\Gamma_{\rm eff}$ values that deviate from the mean by more than $2.5\sigma$ (i.e., $\Gamma_{\rm eff}\la0.5$ or
$\Gamma_{\rm eff}\ga3.1$) come predominantly from these 553 objects. Excluding these 553 objects,
the distribution of $\Gamma_{\rm eff}$ for the 732 objects that are detected in both the soft and hard bands
is shown in Figure~\ref{fig:geff}(b). The best-fit Gaussian profile has a mean of 1.70 and a standard deviation of 0.45.

Based on the above $\Gamma_{\rm eff}$ distributions, we adopted $\Gamma=1.26$ to be the threshold separating \mbox{X-ray} unabsorbed and
potentially \mbox{X-ray} absorbed quasars in this paper (i.e., an object with $\Gamma_{\rm eff}<1.26$ is considered an absorbed quasar),
which corresponds to a negative $1\sigma$ deviation from the means of the Gaussian distributions.
We have verified that adopting a larger $\Gamma_{\rm eff}$ value of 1.4 as the threshold would not affect significantly our results below.
We thus excluded 230 quasars with $\Gamma<1.26$ in our sample when fitting the \mbox{$\alpha_{\rm OX}$--$L_{\rm 2500~{\textup{\AA}}}$}
relation. We utilized the EM (estimate and maximize) algorithm in ASURV to derive the linear regression parameters.
The best-fit relation for the remaining 1595 quasars is
\begin{equation}
\alpha_{\rm OX}=(-0.224 \pm 0.008) \log(L_{\rm 2500~{\textup{\AA}}}) + (5.373 \pm 0.254), 
\label{equ:al1}
\end{equation}
which is shown as the solid black line in Figure~\ref{fig:al_0a}. The \citet{Just2007}
\mbox{$\alpha_{\rm OX}$--$L_{\rm 2500~{\textup{\AA}}}$} relation is shown for comparison.

The slope of the best-fit \mbox{$\alpha_{\rm OX}$--$L_{\rm 2500~{\textup{\AA}}}$} relation is steeper than
that from \citet{Just2007}. We note that our sample A spans about three orders of magnitude in $L_{\rm 2500~{\textup{\AA}}}$
\mbox{[$29.3\la\log(L_{\rm 2500~{\textup{\AA}}})\la32.0$]}, while the \citet{Just2007} sample spans
a much larger $L_{\rm 2500~{\textup{\AA}}}$ range and extends to lower luminosities
\mbox{[$27.5\la\log(L_{\rm 2500~{\textup{\AA}}})\la32.5$]}.
Therefore, the apparent difference between the \mbox{$\alpha_{\rm OX}$--$L_{\rm 2500~{\textup{\AA}}}$} slopes
of this study and \citet{Just2007} may result from the different UV luminosity ranges.
The difference is consistent with the \citet{Steffen2006} interpretation that the
\mbox{$\alpha_{\rm OX}$--$L_{\rm 2500~{\textup{\AA}}}$} slope becomes steeper at higher UV luminosities.
We also note that our \mbox{$\alpha_{\rm OX}$--$L_{\rm 2500~{\textup{\AA}}}$} relation is
consistent with that derived in a recent study by \citet{Timlin2020}.

\subsection{\mbox{X-ray} Weak Quasars and Their Fraction} \label{sec:frac}

\begin{figure}[htb!]
\centerline{
\includegraphics[scale=0.5]{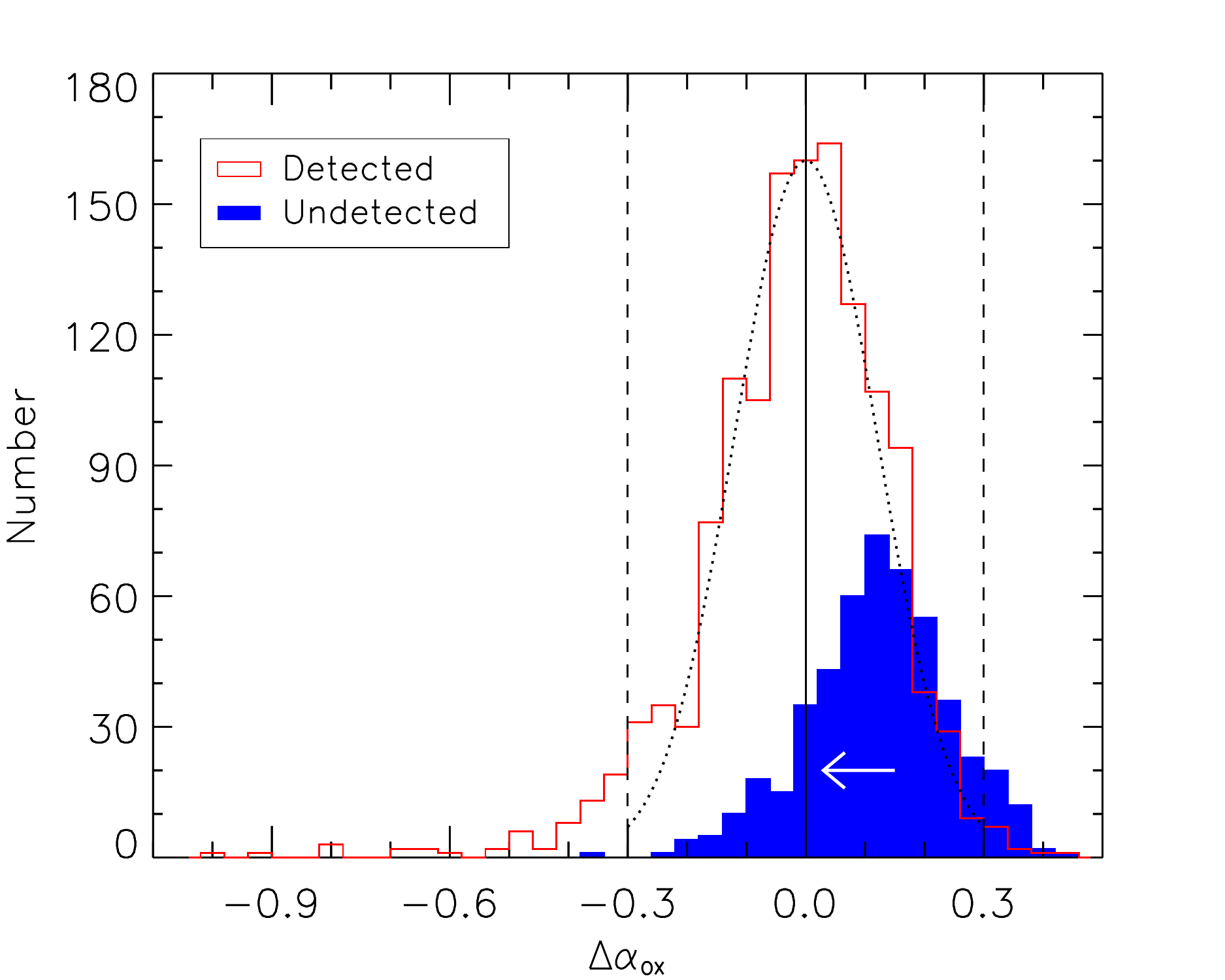}
}
\caption{Distributions of the $\Delta\alpha_{\rm OX}$ values for sample A. The open red histogram represents the distribution
for the \mbox{X-ray} detected objects, while the solid blue histogram represents the distribution of the $3\sigma$ upper limits
on $\Delta\alpha_{\rm OX}$ for the \mbox{X-ray} undetected objects. The arrow indicates that the blue histogram is the distribution
of upper limits. The vertical dashed lines (solid line) indicate $\Delta\alpha_{\rm OX}=\pm0.3$ ($\Delta\alpha_{\rm OX}=0$)
from the best-fit \mbox{$\alpha_{\rm OX}$--$L_{\rm 2500~{\textup{\AA}}}$} relation
(Equation~3; see Section~\ref{sec:al}). The black dotted curve shows a Gaussian profile with a mean value of 0 and
a standard deviation of 0.12; it is not a fit to the $\Delta\alpha_{\rm OX}$ distribution but is drawn to guide the eye.
\label{fig:daox_0a} }
\end{figure}

We calculated the $\Delta\alpha_{\rm OX}$ parameter, defined as the difference between
the measured $\alpha_{\rm OX}$ and that expected from the best-fit \mbox{$\alpha_{\rm OX}$--$L_{\rm 2500~{\textup{\AA}}}$}
relation (Equation~3), $\Delta\alpha_{\rm OX}=\alpha_{\rm OX}-\alpha_{\rm OX}(L_{\rm 2500~{\textup{\AA}}})$.
In order to define the \mbox{X-ray} weak population, we examined the distribution of $\Delta\alpha_{\rm OX}$ values, 
which is shown in Figure~\ref{fig:daox_0a}. Of the 1825 objects in our sample, 65
(including one \mbox{X-ray} undetected object) have $\Delta\alpha_{\rm OX}\leq-0.3$
but only 11 have $\Delta\alpha_{\rm OX}\geq0.3$,
which causes the apparent asymmetry of the $\Delta\alpha_{\rm OX}$ distribution.
The distribution of the \mbox{X-ray} detected objects can be well described by a Gaussian profile plus a negative tail
with $\Delta\alpha_{\rm OX}<-0.3$, indicating the existence of a small population of \mbox{X-ray} weak quasars.
Based on the $\Delta\alpha_{\rm OX}$ distribution, we adopted $\Delta\alpha_{\rm OX} = -0.3$ as the threshold
separating \mbox{X-ray} normal and \mbox{X-ray} weak quasars in this paper (i.e., an object with $\Delta\alpha_{\rm OX}\leq-0.3$
is considered an \mbox{X-ray} weak quasar). We note that the $\Delta\alpha_{\rm OX}$ distribution
for the 1114 $\Gamma\ge1.26$ \mbox{X-ray} detected objects (Section~\ref{sec:al}) has a standard deviation of 0.12.
Thus, $\Delta\alpha_{\rm OX} = -0.3$ corresponds to a negative 2.5$\sigma$ deviation from the zero point.

\begin{figure}[htb!]
\centerline{
\includegraphics[scale=0.5]{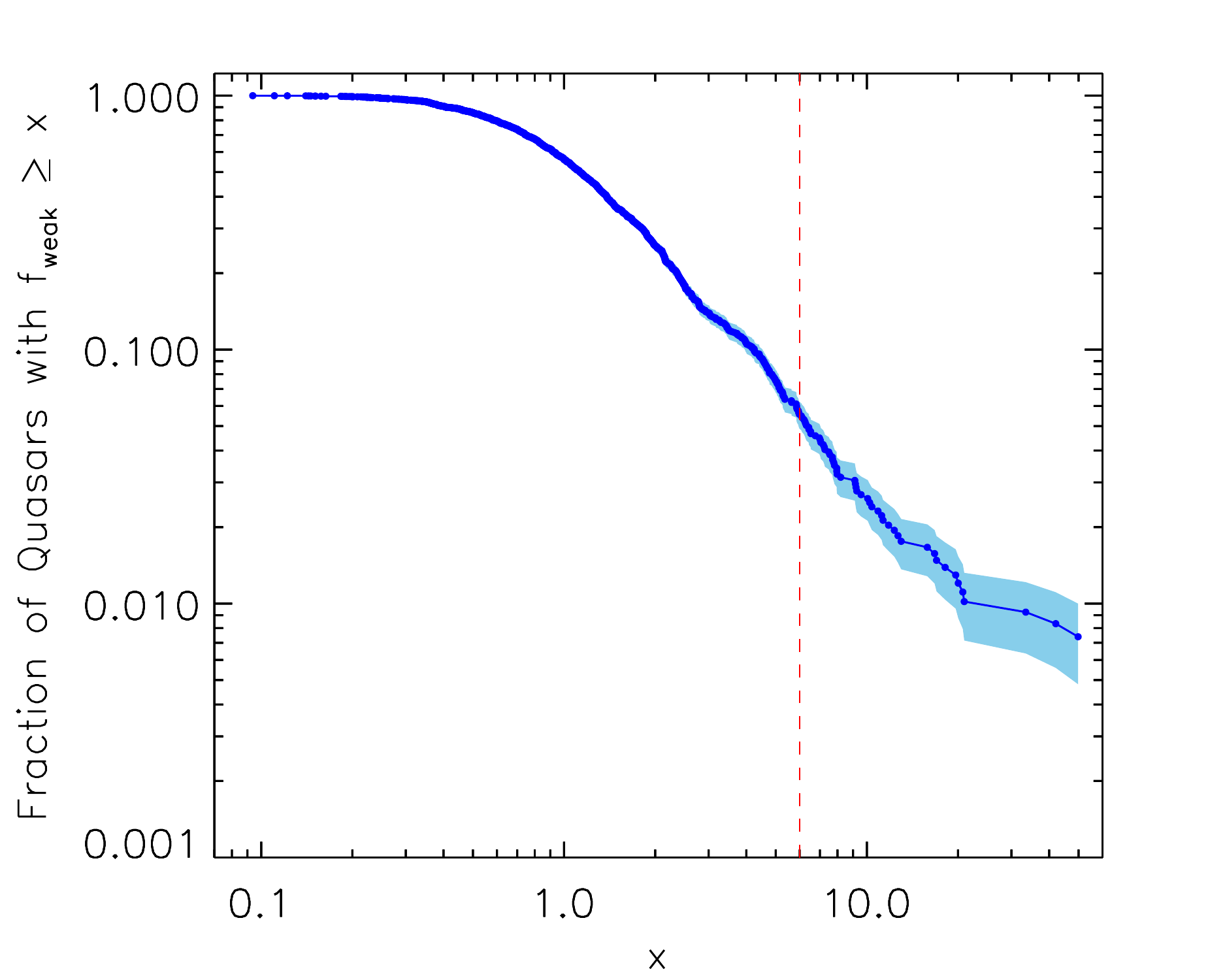}
}
\caption{Fraction of quasars with $f_{\rm weak}\ge x$ vs. $x$.
The shaded region indicates the $1\sigma$ confidence intervals calculated following the method of \citet{Avni1980}.
The vertical line shows $f_{\rm weak}\ge6$ (\mbox{$\Delta\alpha_{\rm OX}\le-0.3$}),
which is the threshold separating \mbox{X-ray} normal and \mbox{X-ray} weak quasars.
\label{fig:frac_2a} }
\end{figure}

\citet{Gibson2008a} have measured upper limits on the fractions of quasars that are \mbox{X-ray} weak by given factors. In this study,
as the number of \mbox{X-ray} quasars has been increased significantly, we can better constrain the \mbox{X-ray} weak fractions.
Considering that the \mbox{X-ray} undetected quasars have only upper limit constraints on their \mbox{X-ray} weakness factors,
we utilized the Kaplan-Meier estimator provided in the ASURV package, which works with censored data,
to derive the best estimates and uncertainties for these fractions. Figure~\ref{fig:frac_2a} shows the fraction of quasars
with $f_{\rm weak}\ge x$ versus $x$. The \mbox{X-ray} weakness factor $f_{\rm weak}$ was measured from
$\Delta\alpha_{\rm OX}$ as $f_{\rm weak}=10^{-\Delta\alpha_{\rm OX}/0.3838}$, and the \mbox{X-ray} weak threshold $\Delta\alpha_{\rm OX}\le-0.3$
corresponds to an \mbox{X-ray} weakness factor of $f_{\rm weak}\ge6.0$. We found the fraction of \mbox{X-ray} weak quasars to be 
$5.8\pm0.7\%$; the $1\sigma$ uncertainties were calculated following the method of \citet{Avni1980}.
Additionally, the fractions of quasars that are \mbox{X-ray} weak by factors of \mbox{$f_{\rm weak}\geq10$}
(\mbox{$\Delta\alpha_{\rm OX}\leq-0.38$}), \mbox{$f_{\rm weak}\geq20$} (\mbox{$\Delta\alpha_{\rm OX}\leq-0.50$}),
and \mbox{$f_{\rm weak}\geq50$} (\mbox{$\Delta\alpha_{\rm OX}\leq-0.65$}) are \mbox{$2.7\pm0.5\%$}, \mbox{$1.3\pm0.3\%$},
and \mbox{$0.74\pm0.26\%$}, respectively.

\subsection{Fractions of \mbox{X-ray} Weak Quasars among Two Subsamples} \label{sec:frac_BC}

We note that the Kaplan-Meier estimator used in Section~\ref{sec:frac} assumes that the intrinsic distribution of the censored data
is the same as that of the measured data \citep{FN1985}. The assumption is sensible if the \mbox{X-ray} nondetections are simply caused
by shallower Chandra coverage, but it is not valid if many of the \mbox{X-ray} undetected quasars are indeed more \mbox{X-ray} weak
(having a more negative $\Delta\alpha_{\rm OX}$ distribution) than the \mbox{X-ray} detected quasars. 
Given Figure~\ref{fig:ta}, it appears that the \mbox{X-ray} undetected quasars in general have shallower Chandra coverage than
the \mbox{X-ray} detected quasars, but it is not sufficient to justify the assumption above, especially when we are interested in the exceptionally
\mbox{X-ray} weak quasars. Therefore, in this subsection, we constructed, without direct reference to \mbox{X-ray} source properties, 
two subsamples of quasars out of sample A that do not have censored $\Delta\alpha_{\rm OX}$ values
(all \mbox{X-ray} detected), and then we compared the fractions of \mbox{X-ray} weak quasars among these subsamples to that in sample A.

The principle to select quasars that are likely \mbox{X-ray} detected is to limit the sample to bright quasars and to also limit the
\mbox{X-ray} observations to those with good sensitivity (relatively large exposures and small off-axis angles; e.g., Figure~\ref{fig:ta}).
Because the DR7 and DR10 quasar catalogs have significantly different depths, we cannot select bright quasars from these two catalogs
using a uniform brightness criterion. Thus, we constructed a subsample of quasars from the DR7 and DR10 catalogs, respectively.
The detailed criteria are as follows:
\begin{enumerate} 
\item
{\it i}-band magnitude $m_{i}<19.6$ ($m_{i}<21.1$) for DR7 (DR10) quasars,
\item
effective exposure time $>$2.5~ks ($>$6.3~ks) and off-axis angle $<$9.8\arcmin ($<$8.2\arcmin)
for DR7 (DR10) quasars. We adopted more stringent exposure-time
and off-axis angle criteria for DR10 quasars, since they are on average optically fainter than DR7 quasars.
\end{enumerate}
These two criteria resulted in a subsample of 218 (208) quasars in DR7 (DR10), and we refer to this subsample as sample B (sample C).
The selections of samples B and C are also listed in Table~1. Quasars in sample B or sample C are all \mbox{X-ray} detected, and thus we can
compute the fraction of \mbox{X-ray} weak quasars among each subsample directly, without employing the Kaplan-Meier estimator. 
We stress that we do not require \mbox{X-ray} detections in the construction of samples B and C, which are simply the outcomes of the
appropriate brightness and \mbox{X-ray} sensitivity criteria listed above.

\begin{figure*}[htb!]
\centerline{
\includegraphics[scale=0.5]{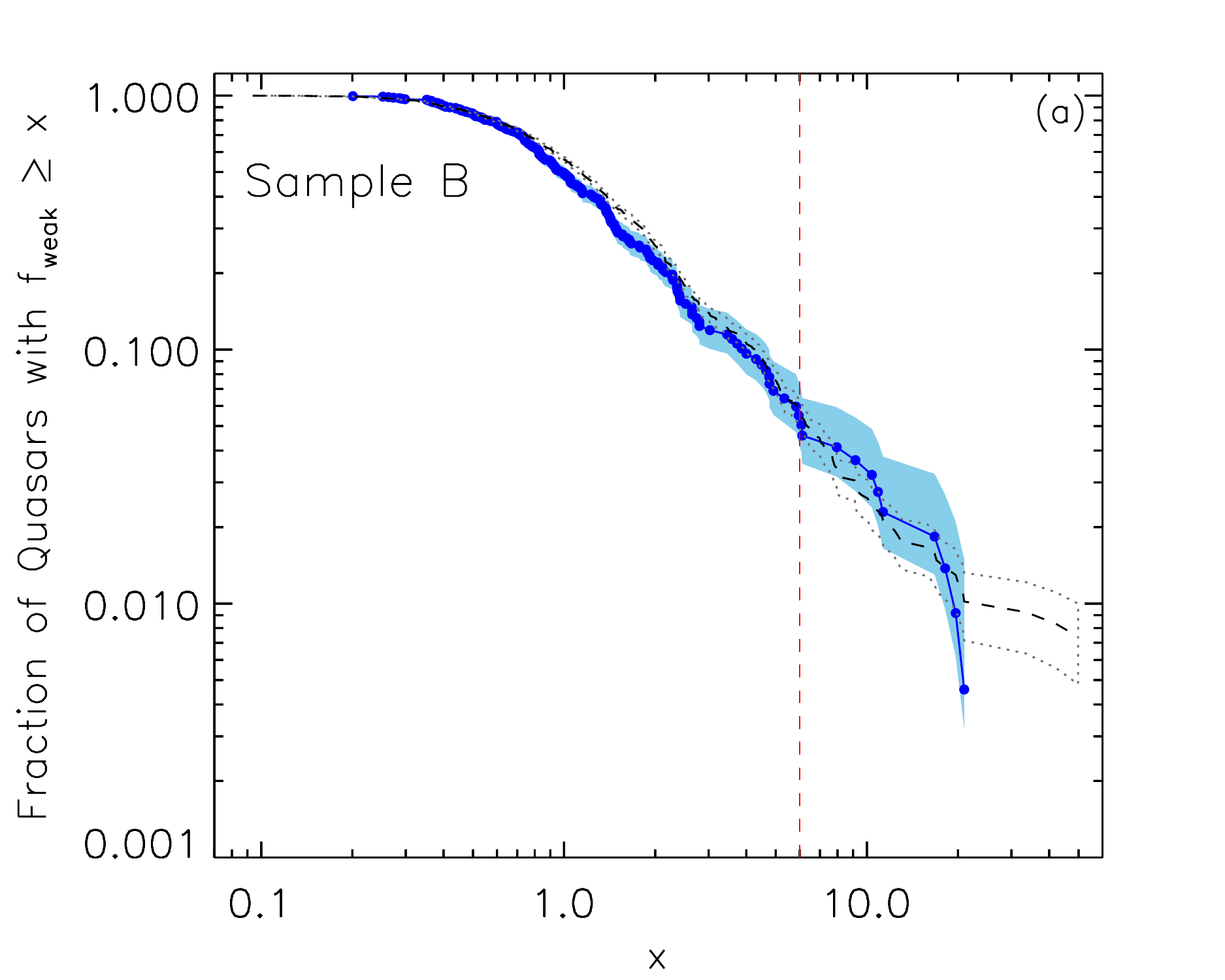}
\includegraphics[scale=0.5]{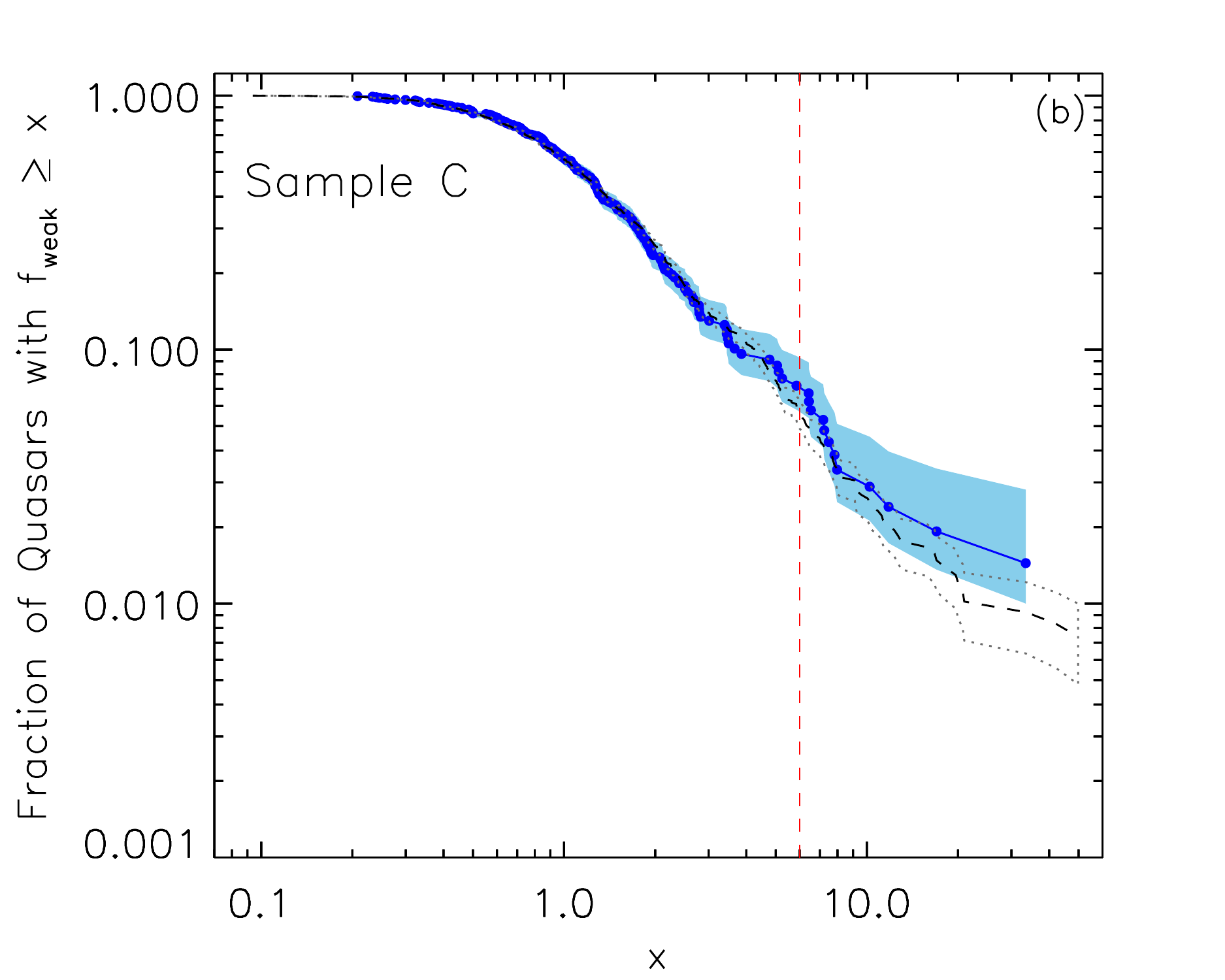}
}
\caption{Fraction of quasars with $f_{\rm weak}\ge x$ vs. $x$ for (a) sample B and (b) sample C.
The shaded region indicates the $1\sigma$ binomial confidence intervals calculated following the method of
\citet{Cameron2011}. The vertical line in each panel shows $f_{\rm weak}\ge6$ (\mbox{$\Delta\alpha_{\rm OX}\le-0.3$}),
which is the threshold separating \mbox{X-ray} normal and \mbox{X-ray} weak quasars.
The \mbox{X-ray} weak fraction (black dashed curve) and associated $1\sigma$ confidence intervals (black dotted curves)
for sample A are plotted in both panels for comparison.
\label{fig:frac_2} }
\end{figure*}

Given our \mbox{X-ray} weak definition of $\Delta\alpha_{\rm OX}\leq-0.3$, sample B contains 12 \mbox{X-ray} weak quasars,
corresponding to an \mbox{X-ray} weak fraction of $ 5.5_{- 1.2}^{+ 2.0}\%$ (12/218); the $1\sigma$ binomial confidence
intervals were calculated following the method of \citet{Cameron2011}. Figure~\ref{fig:frac_2}(a) shows the fraction of quasars with
$f_{\rm weak}\ge x$ versus $x$ in sample B. The fraction of quasars that are \mbox{X-ray} weak by a factor of \mbox{$f_{\rm weak}\geq10$}
(\mbox{$\Delta\alpha_{\rm OX}\leq-0.38$}) in sample B is $ 3.2_{- 0.8}^{+ 1.7}\%$ (7/218). In sample C, there are 14
\mbox{X-ray} weak quasars, corresponding to an \mbox{X-ray} weak fraction of $ 6.7_{- 1.4}^{+ 2.2}\%$ (14/208).
Figure~\ref{fig:frac_2}(b) shows the fraction of quasars with $f_{\rm weak}\ge x$ versus $x$ in sample C.
The \mbox{$f_{\rm weak}\geq10$} fraction in sample C is $ 2.9_{- 0.8}^{+ 1.7}\%$ (6/208).
The \mbox{X-ray} weak fraction curve for sample A is also plotted in Figure~\ref{fig:frac_2}, showing general consistency with the curve
for sample B or sample C.

We note that the fractions of \mbox{X-ray} weak quasars among samples B and C are consistent within the uncertainties.
Samples B and C probe slightly different quasar populations in terms of their optical brightness and \mbox{X-ray} coverage,
and the fractions of \mbox{X-ray} weak quasars among them are not necessarily the same, but our current data is not sufficient to
determine if the fraction has any luminosity dependence. In order to increase the sample size and reduce the fraction uncertainties,
we combine samples B and C (sample B+C) in our following analysis. In this combined sample, the \mbox{X-ray} weak and
\mbox{$f_{\rm weak}\geq10$} fractions become $ 6.1_{- 1.0}^{+ 1.4}\%$ (26/426) and $ 3.1_{- 0.6}^{+ 1.1}\%$ (13/426),
respectively. Comparing these fractions to those in sample A, we found that they are consistent within the uncertainties.
The Kaplan-Meier estimator appears to provide a reasonable estimate of the fraction of \mbox{X-ray} weak quasars in sample A, and
the difference between the quasar populations in sample A and sample B+C does not appear to affect significantly the \mbox{X-ray} weak fractions.   

\section{Discussion} \label{sec:Dis}

\subsection{Fractions of \mbox{X-ray} Weak Quasars} \label{sec:fxw}

Our study found that a small fraction of optically selected, non-BAL, type 1 quasars do show weak \mbox{X-ray} emission.
Based on our sample A of 1825 quasars, 1344 (\mbox{$74\pm1\%$}) of which are \mbox{X-ray} detected, the fraction of \mbox{X-ray} weak
($f_{\rm weak}\ge6$) quasars is \mbox{$5.8\pm0.7\%$} and the fraction of $f_{\rm weak}\ge10$ quasars is \mbox{$2.7\pm0.5\%$}.
Based on our subsample B+C of 426 quasars that are composed of \mbox{X-ray} detections only, the fraction of \mbox{X-ray} weak quasars is
$ 6.1_{- 1.0}^{+ 1.4}\%$ and the fraction of $f_{\rm weak}\ge10$ quasars is $ 3.1_{- 0.6}^{+ 1.1}\%$. These small but non-negligible
fractions of \mbox{X-ray} weak quasars challenge the ubiquity of quasar \mbox{X-ray} emission, and they also require additional physical mechanisms internal
or external to the corona that can suppress the observed \mbox{X-ray} emission; we discuss some possible causes for the \mbox{X-ray} weak quasars in 
Section~\ref{sec:nxw} below.
 
Our current investigation was motivated by the \citet{Gibson2008a} study that used a much smaller sample of quasars. Based on the sample B
of \citet{Gibson2008a}, which is similar to our sample B+C here composed of \mbox{X-ray} detections only but has only 139 quasars, the
fraction of $f_{\rm weak}\ge10$ quasars is considered to be $\lesssim2\%$. Their result is consistent with ours here considering
the uncertainties of both studies.

Recently, a few studies have found much larger fractions of \mbox{X-ray} weak quasars, albeit among samples with very limited sizes.
For example, \citet{Nardini2019} found seven \mbox{X-ray} weak quasars among 29 very luminous RQ quasars at $z\approx3.0\textrm{--}3.3$,
corresponding to an \mbox{X-ray} weak fraction of 24\%. In addition, \citet{Zappacosta2020} found $4\textrm{--}5$ \mbox{X-ray} weak quasars among 13 hyper
luminous RQ quasars at $z\approx2\textrm{--}4$, corresponding to an \mbox{X-ray} weak fraction of $\approx30\textrm{--}40\%$. The definitions of
\mbox{X-ray} weak quasars in these studies are slightly different from our adopted $\Delta\alpha_{\rm OX}\leq-0.3$ here, but adopting
their definitions would not change significantly the fractions of \mbox{X-ray} weak quasars in our samples and they are still only about $5\textrm{--}7\%$.
There are several factors that may contribute to the discrepancy between the fractions of \mbox{X-ray} weak quasars, and detailed comparison
between their result and the \citet{Gibson2008a} sample B result was also presented in Section 5.7 of \citet{Nardini2019}. We list two
factors below that may be responsible for most of the discrepancy.
\begin{enumerate}
\item
The fractions among small quasar samples are vulnerable to contamination from a few \mbox{X-ray} weak BAL quasars. There may be missed \mbox{X-ray}
weak BAL quasars in the \citet{Nardini2019} and \citet{Zappacosta2020} samples (e.g., Section 5.8 and Appendix B of \citealt{Nardini2019}),
and even one such quasar in the sample would bias the fraction significantly.
\item
The fraction of \mbox{X-ray} weak quasars likely has a luminosity dependence, and it becomes larger among more luminous quasars.
The main difference between the \citet{Nardini2019} and \citet{Zappacosta2020} samples and our sample is the quasar luminosity.
The 29 quasars in \citet{Nardini2019} and 13 quasars in \citet{Zappacosta2020} have a UV luminosity range of
\mbox{$31.8\la\log(L_{\rm 2500~{\textup{\AA}}})\la32.5$} except one object, while our sample quasars have
\mbox{$29.3\la\log(L_{\rm 2500~{\textup{\AA}}})\la32.0$}.
We do not have a sizable number of very luminous quasars for direct comparison,
but if we limit our sample to the 30 most luminous quasars in sample A, which have
\mbox{$31.4\la\log(L_{\rm 2500~{\textup{\AA}}})\la32.0$} and a median $\log(L_{\rm 2500~{\textup{\AA}}})$ value of 31.5,
the fraction of \mbox{X-ray} weak quasars in this high-luminosity sample becomes $16\pm8\%$,
much larger compared to the fraction in the full sample. If we limit our sample to the 50 most luminous quasars in sample A,
which have \mbox{$31.3\la\log(L_{\rm 2500~{\textup{\AA}}})\la32.0$} and a median $\log(L_{\rm 2500~{\textup{\AA}}})$ value of 31.4,
the \mbox{X-ray} weak fraction becomes $15\pm6\%$. Such a luminosity dependence for the fraction of \mbox{X-ray} weak quasars
is in general agreement with our interpretations of \mbox{X-ray} weak quasars as WLQs (e.g., \citealt{Luo2015,Ni2018}; Section~\ref{sec:WLQ})
or due to extreme \mbox{X-ray} variability (e.g., \citealt{Liu2019,Ni2020}; Section~\ref{sec:unc}), where quasars accreting at high Eddington ratios may
have \mbox{X-ray} absorption from a geometrically thick inner accretion disk or its associated outflow; the fraction of such quasars is likely higher
among more luminous samples. The relevance of the enhanced fraction of \mbox{X-ray} weak quasars to WLQs is also noted in both
\citet{Nardini2019} and \citet{Zappacosta2020}.
We also note that if we divide our sample A into a high-luminosity subsample and a low-luminosity subsample at the median luminosity of
$\log(L_{\rm 2500~{\textup{\AA}}})=30.45$, the fractions of \mbox{X-ray} weak quasars among the two subsamples are $6.2\pm1.0\%$ and
$5.4\pm1.0\%$, respectively. Such a small difference indicates that the enhanced fraction of \mbox{X-ray} weak quasars
(probably related to the enhanced fraction of WLQs) is mostly evident among very luminous quasars.
\end{enumerate}
Given the above comparisons, we stress that the fractions of \mbox{X-ray} weak quasars derived in this study are mostly applicable to quasar
samples sharing similar properties to our SDSS quasars here (Table~1), and extra caution is needed if quoting the fractions
for quasars with substantially different luminosities.

Finally, we constrain the fraction of intrinsically \mbox{X-ray} weak quasars. In sample B+C, there are nine \mbox{X-ray} weak
objects with $\Gamma_{\rm eff}\ge1.26$ (including three with $\Gamma_{\rm eff}$ set to 1.8), which we consider to be candidates for
intrinsically \mbox{X-ray} weak quasars. Thus, the fraction of intrinsically \mbox{X-ray} weak quasars within the non-BAL quasar
population is $ 1.4_{- 0.4}^{+ 0.8}\%\textrm{--} 2.1_{- 0.5}^{+ 0.9}\%$ ($6/426\textrm{--}9/426$).
This fraction would be $ 1.2_{- 0.3}^{+ 0.8}\%\textrm{--} 1.9_{- 0.5}^{+ 0.9}\%$ ($5/426\textrm{--}8/426$)
if we consider \mbox{X-ray} weak objects with $\Gamma_{\rm eff}\ge1.4$ to be candidates for intrinsically \mbox{X-ray} weak quasars.
Based on the \mbox{X-ray} properties of 35 Large Bright Quasar Survey (LBQS) BAL quasars, \citet{Liu2018} estimated the fraction
of intrinsically \mbox{X-ray} weak quasars within the BAL quasar population to be $ 5.7_{- 1.9}^{+ 6.7}\%\textrm{--}23_{- 6}^{+ 8}\%$
($2/35\textrm{--}8/35$), which is significantly higher than the fraction within the non-BAL quasar population derived here.
These results indicate that intrinsically \mbox{X-ray} weak quasars may be preferentially found in BAL quasars \citep[e.g.,][]{Luo2014,Liu2018},
and intrinsically \mbox{X-ray} weak non-BAL quasars like PHL~1811 are extremely rare.

\subsection{Nature of \mbox{X-ray} Weak Quasars} \label{sec:nxw}

\begin{deluxetable*}{l|ccccc|cccc}
\tabletypesize{\normalsize}
\tablecaption{Numbers of Potentially Absorbed and Unabsorbed Quasars in the \mbox{X-ray} Weak Populations}
\tablehead{
\multicolumn{1}{c|}{}    &
\multicolumn{5}{c|}{Sample A} &
\multicolumn{4}{c}{Sample B+C}  \\
\multicolumn{1}{c|}{\mbox{X-ray} Weak}    &
\colhead{Undetected}         &
\multicolumn{4}{c|}{Detected} &
\multicolumn{4}{c}{Detected}  \\
\multicolumn{1}{c|}{Population}    &
\colhead{All}    &
\colhead{All}    &
\colhead{Abs.}    &
\colhead{Unabs.}    &
\multicolumn{1}{c|}{Unclear}   &
\colhead{All}    &
\colhead{Abs.}    &
\colhead{Unabs.}    &
\colhead{Unclear}
}
\startdata
WLQ\tablenotemark{a}         & $           1$ & $           9$ & $           6$ & $           2$ & $           1$ & $
           7$ & $           4$ & $           2$ & $           1$ \\
Red quasar\tablenotemark{b}  & $           1$ & $          13$ & $          10$ & $           1$ & $           2$ & $
           8$ & $           6$ & $           0$ & $           2$ \\
\ \ \ Red WLQ                & $           1$ & $           3$ & $           2$ & $           0$ & $           1$ & $
           3$ & $           2$ & $           0$ & $           1$ \\
\ \ \ Red non-WLQ            & $           0$ & $          10$ & $           8$ & $           1$ & $           1$ & $
           5$ & $           4$ & $           0$ & $           1$ \\
Unclassified\tablenotemark{c}& $           0$ & $          45$ & $          37$ & $           7$ & $           1$ & $
          14$ & $           9$ & $           4$ & $           1$ \\
Total                        & $           1$ & $          64$ & $          51$ & $          10$ & $           3$ & $
          26$ & $          17$ & $           6$ & $           3$ \\
\enddata
\tablecomments{We adopted $\Delta\alpha_{\rm OX} = -0.3$ to be the threshold separating \mbox{X-ray} normal and \mbox{X-ray}
weak quasars (Section~\ref{sec:frac}).
All the undetected \mbox{X-ray} weak quasars in sample A have fixed \mbox{X-ray} photon indices of $\Gamma_{\rm eff}=1.8$;
the numbers of detected \mbox{X-ray} weak quasars with fixed $\Gamma_{\rm eff}$ values of 1.8 (see Section~\ref{sec:X})
are listed in the ``Unclear'' columns.}
\tablenotetext{\rm a}{We adopted \iona{C}{iv} REW $\rm {=16~\AA}$ to be the threshold separating WLQs and normal quasars (Section~\ref{sec:WLQ}).}
\tablenotetext{\rm b}{We adopted $\Delta(g-i)=0.2$ to be the threshold separating red and normal quasars (Section~\ref{sec:red}).}
\tablenotetext{\rm c}{We consider an \mbox{X-ray} weak quasar to be in the unclassified population if it is neither a WLQ nor a red quasar
(Section~\ref{sec:unc}).}
\label{tbl:weak}
\end{deluxetable*}

We discuss the possible nature of the \mbox{X-ray} weak quasars found in this study. We focus on the 26 \mbox{X-ray} weak
($\Delta\alpha_{\rm OX}\leq-0.3$) quasars in sample B+C (12 in sample B and 14 in sample C), but we also
present the overall properties for the \mbox{X-ray} weak quasars in sample A. We consider that there are three types of \mbox{X-ray} weak
quasars based on the optical spectral features: WLQs, red quasars, and unclassified \mbox{X-ray} weak quasars. A breakdown
of the \mbox{X-ray} weak quasars in samples A and B+C is listed in Table~\ref{tbl:weak}.
\linebreak

\subsubsection{\mbox{X-ray} Weak WLQs} \label{sec:WLQ}

With a large fraction ($\ga50\%$) of quasars showing weak \mbox{X-ray} emission, WLQs represent
one population of \mbox{X-ray} weak quasars \citep[e.g.,][]{Wu2012,Luo2015,Ni2018,Timlin2020}.
\mbox{X-ray} weak WLQs were also found to have average small effective photon index, suggestive
of \mbox{X-ray} absorption \citep[e.g.,][]{Luo2015}.
Their \mbox{X-ray} emission may be absorbed by ``shielding gas'' \citep[e.g.,][]{Wu2011,Wu2012}
or a geometrically thick inner accretion disk \citep[e.g.,][]{Luo2015,Ni2018,Ni2020}.

There is no uniform definition for WLQs. For high-redshift ($z > 2.200$) quasars, WLQs are generally selected using the REW of the
\lyanv\ emission lines. \citet{Shemmer2009} defined WLQs to be quasars with \lyanv\ REW $\rm {<10~\AA}$; \citet{DS2009}
defined WLQs to be those with \lyanv\ REW $\rm {<15.4~\AA}$ ($>3\sigma$ deviation from the mean of the log-normal distribution).
For low-redshift ($z\le2.200$) quasars, which do not have \lyanv\ coverage in their optical spectra,
the definition of WLQs is commonly based on other strong emission lines
(e.g., \iona{C}{iv}, \iona{C}{iii]}, \iona{Mg}{ii}, and H$\beta$). \citet{Wu2012} and \citet{Luo2015} selected
WLQs from the \citet{Plotkin2010b} catalog, which has REW $\rm {\la5~\AA}$ for all emission features.
Recently, \citet{Ni2018} called quasars with \mbox{\iona{C}{iv} REW $\rm {<7.0~\AA}$} ``extreme'' WLQs, and
quasars with \mbox{\iona{C}{iv} REW $\rm {=7.0\textrm{--}15.5~\AA}$} ``bridge'' WLQs, which have emission-line features
in between those of extreme WLQs and typical quasars.

\begin{figure*}[htb!]
\centerline{
\includegraphics[scale=1.0]{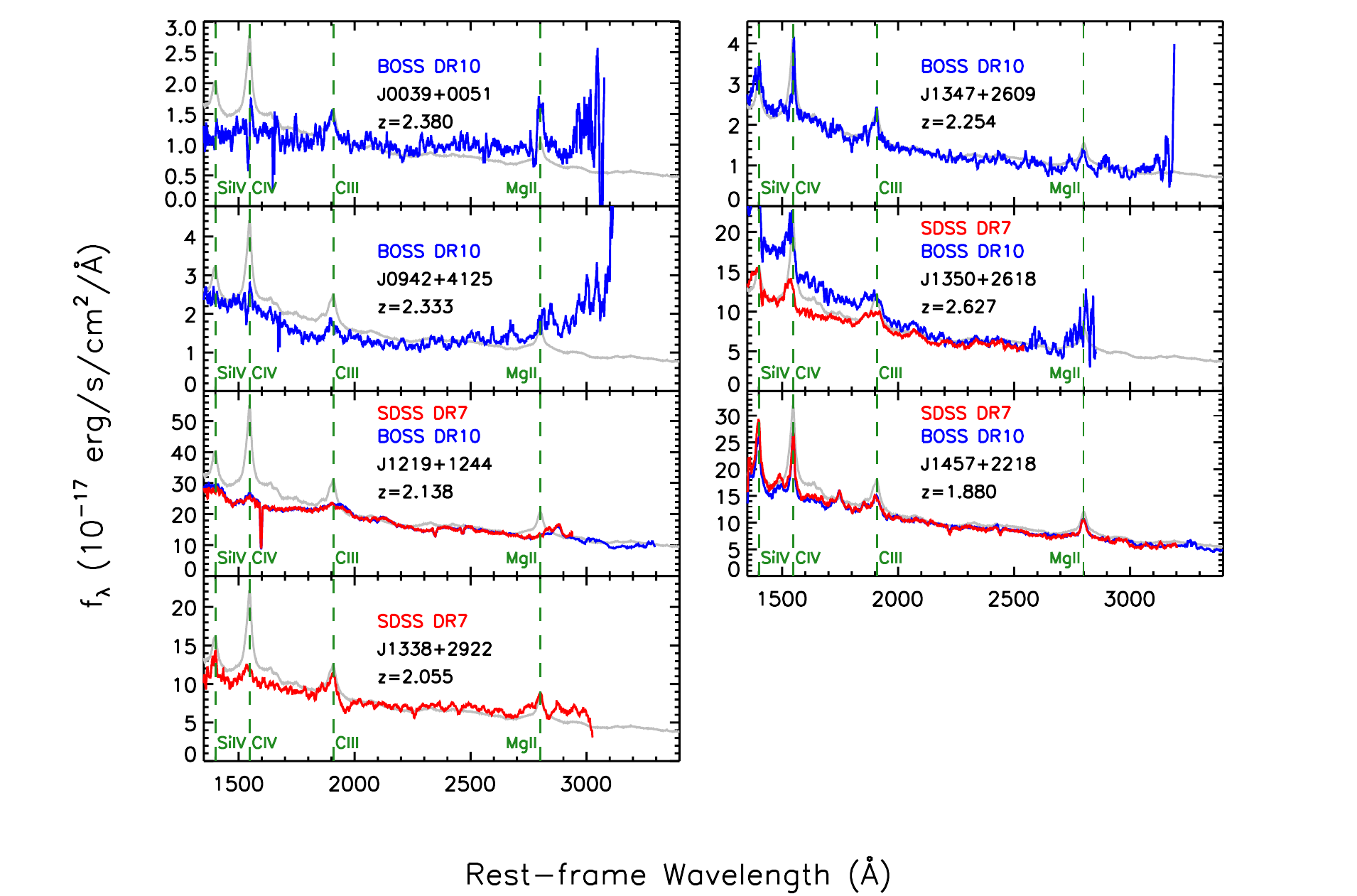}
}
\caption{SDSS optical/UV spectra (smoothed by a 20 pixel boxcar) for the seven \mbox{X-ray} weak WLQs (red curves
for SDSS DR7 spectra and blue curves for BOSS DR10 spectra). The gray curve in each panel shows the SDSS composite quasar
spectrum \citep{Vanden Berk2001} for comparison. SDSS Data Release number, name, and redshift for each object are listed.
\label{fig:sp_wlq} }
\end{figure*}

The \iona{C}{iv} REW values of our 26 \mbox{X-ray} weak quasars in sample B+C range from 6.6 to $\rm {56.8~\AA}$,
with a median value of $\rm {24.4~\AA}$ and a mean value of $\rm {25.6~\AA}$. We adopted \iona{C}{iv} REW $\rm {=16~\AA}$
to be the threshold separating WLQs and normal quasars in this paper. Our more inclusive definition
\iona{C}{iv} REW $\rm {<16~\AA}$ for WLQs corresponds to a $>2.0\sigma$ deviation from the mean of the log-normal distribution
for SDSS quasars, where the distribution of \iona{C}{iv} REW shows an apparent tail toward small values
(see Section 3.2 of \citealt{Wu2012}). Since all of our sample objects have \iona{C}{iv} coverage,
we consider this criterion appropriate for this study. For any sample object that has both SDSS DR7 and
BOSS DR10 spectra, we consider it a WLQ if either spectrum has {\rm \iona{C}{iv}} REW $\rm {<16~\AA}$. 
Given this criterion, sample B+C contains seven \mbox{X-ray} weak WLQs.
Figure~\ref{fig:sp_wlq} shows the spectra of these seven quasars. We note that three of them
(J1219$+$1244, J1350$+$2618, and J1457$+$2218) have both SDSS DR7 and BOSS DR10 spectra. For
J1219$+$1244 and J1350$+$2618, all their spectra satisfy the WLQ criterion, while for J1457$+$2218, only its BOSS DR10
spectrum does.

Given our WLQ definition of \iona{C}{iv} REW $\rm {<16~\AA}$, sample B+C contains 20 WLQs, seven of which are
\mbox{X-ray} weak, corresponding to an \mbox{X-ray} weak fraction of $35_{- 9}^{+12}\%$.
The seven \mbox{X-ray} weak WLQs have a comparable \iona{C}{iv} REW distribution to the other 13 \mbox{X-ray} normal WLQs;
e.g., the median \iona{C}{iv} REWs are 11.9 and $\rm {9.2~\AA}$, respectively.
By contrast, the fraction of \mbox{X-ray} weak quasars is only $ 4.7_{- 0.8}^{+ 1.3}\%$ (19/406) among the non-WLQs in sample B+C.
Thus, WLQs show significantly higher \mbox{X-ray} weak fraction when compared to non-WLQs.
These results confirm previous findings that WLQs represent one population of \mbox{X-ray} weak quasars
\citep[e.g.,][]{Wu2012,Luo2015,Ni2018}.

\citet{Ni2018} investigated a sample of 32 WLQs selected mainly from the \citet{Shen2011} and \citet{Plotkin2010b} catalogs.
They adopted a definition of \iona{C}{iv} REW $\rm {<15.5~\AA}$ for ``extreme''+``bridge'' WLQs, which is similar to
our WLQ definition, and the fraction of \mbox{X-ray} weak objects within their WLQ sample is $44_{- 8}^{+ 9}\%$ (14/32).
We note that \citet{Ni2018} adopted $\Delta\alpha_{\rm OX}<-0.2$ to define \mbox{X-ray} weak quasars. 
If we use this same definition, sample B+C would contain ten \mbox{X-ray} weak WLQs and the \mbox{X-ray} weak fraction
for WLQs would be $50\pm11\%$, which is in agreement with \citet{Ni2018}.

\citet{Wu2011} proposed a ``shielding gas'' scenario to unify \mbox{X-ray} normal and \mbox{X-ray} weak WLQs. In this scenario,
a small fraction of the quasar population has high-ionization shielding gas lying between the SMBH and the
broad emission-line region (BELR). With high column density and large BELR covering factor, the shielding gas
is able to prevent most, if not all, \mbox{X-ray} emission and other ionizing continuum from reaching the BELR. If such a quasar
is viewed through the shielding gas, an \mbox{X-ray} weak WLQ is seen; if it is viewed from other orientations,
an \mbox{X-ray} normal WLQ is seen. The average hard \mbox{X-ray} spectrum via stacking analyses, which is suggestive of
\mbox{X-ray} absorption, further supports the shielding-gas scenario \citep{Luo2015}.

The $\Gamma_{\rm eff}$ values of the seven \mbox{X-ray} weak WLQs in sample B+C are listed in Table~\ref{tbl:WLQ}. Given our
$\Gamma_{\rm eff}<1.26$ criterion for being potentially \mbox{X-ray} absorbed (Section~\ref{sec:al}), four of the seven
\mbox{X-ray} weak WLQs are potentially \mbox{X-ray} absorbed, with their $\Gamma_{\rm eff}$ values ranging from $-0.8$ to 1.1. 
The small effective photon indices for these four \mbox{X-ray} absorbed WLQs are consistent with the shielding-gas scenario
\citep[e.g.,][]{Wu2011,Wu2012}. In addition, there are two apparently unabsorbed \mbox{X-ray} weak WLQs, J1350$+$2618 and J1457$+$2218, which
have $\Gamma_{\rm eff}=2.0_{-1.0}^{+1.9}$ and $1.6_{-0.2}^{+0.3}$, respectively. These two are good candidates for
intrinsically \mbox{X-ray} weak quasars. One of them, J1457$+$2218, is detected in all three \mbox{X-ray} bands. With
70 photons in the broad band, it is the only object that has $>50$ broad-band photons in sample B+C.
We present \mbox{X-ray} spectral analysis of J1457$+$2218 in Appendix~\ref{sec:Xspec}.
Another quasar, J1219$+$1244, is undetected in both the soft and hard bands and has a fixed photon index of
$\Gamma_{\rm eff}=1.8$, and thus it is unclear whether this object is \mbox{X-ray} absorbed.

However, we note that, except for J1457$+$2218, the other six \mbox{X-ray} weak WLQs have only 2--22 photons in the broad band,
leading to substantial uncertainties of the effective photon indices $\Gamma_{\rm eff}$ (see Table~\ref{tbl:WLQ}).
We also note that four \mbox{X-ray} weak WLQs (J0039$+$0051, J1219$+$1244, J1347$+$2609, and J1350$+$2618) are mini-BAL quasars
(see Section~\ref{sec:unc} below), and their \mbox{X-ray} weakness might also be related to their mini-BAL features, although
mini-BALs do not necessarily lead to weak \mbox{X-ray} emission.

\begin{deluxetable}{lcc}
\tabletypesize{\normalsize}
\tablecaption{Broad-band Counts and $\Gamma_{\rm eff}$ Values for the \mbox{X-ray} Weak WLQs in Sample B+C}
\tablehead{
\colhead{}  &
\colhead{Broad-band} &
\colhead{} \\
\colhead{Object}  &
\colhead{Counts} &
\colhead{$\Gamma_{\rm eff}$}
}
\startdata
J$003927.13+005152.8$ & $   2$ & $-0.8_{-3.3}^{+1.4}$ \\
J$094200.66+412545.3$ & $   7$ & $-0.4\pm1.0$ \\
J$121946.20+124454.1$\tablenotemark{a} & $   2$ & $ 1.8$ \\
J$133820.04+292206.1$ & $  17$ & $ 1.1_{-1.0}^{+0.9}$ \\
J$134745.49+260940.1$ & $  22$ & $ 1.0\pm0.4$ \\
J$135058.12+261855.2$ & $  10$ & $ 2.0_{-1.0}^{+1.9}$ \\
J$145710.80+221844.8$ & $  70$ & $ 1.6_{-0.2}^{+0.3}$ \\
\enddata
\tablecomments{We adopted \iona{C}{iv} REW $\rm {=16~\AA}$ for the threshold separating WLQs and normal quasars in this paper.}
\tablenotetext{\rm a}{J$1219+1244$ is undetected in both the soft and hard bands and has a fixed \mbox{X-ray} photon index of \linebreak
$\Gamma_{\rm eff}=1.8$.}
\label{tbl:WLQ}
\end{deluxetable}

Based on the \citet{Wu2011} shielding-gas scenario, \citet{Luo2015} proposed that the shielding gas may be
a geometrically thick inner accretion disk. When a quasar is accreting at a high Eddington ratio ($L_{\rm Bol}/L_{\rm Edd}\ga0.3$),
its inner accretion disk may become significantly puffed up \citep[e.g.,][]{KB1999,Blaes2001,LD2011,SN2012,NT2014}. 
The puffed-up disk can block nuclear \mbox{X-ray}s and other ionizing photons from reaching an equatorial BELR. Thus, an \mbox{X-ray} weak
WLQ can be seen when such a quasar is viewed through the thick inner disk.

To test the \citet{Luo2015} model, we estimated the $L_{\rm Bol}/L_{\rm Edd}$ values of our sample objects from
their bolometric luminosities and SMBH masses. We measured SMBH masses using the \iona{Mg}{ii} \citep{VO09} or \iona{C}{iv}
\citep{VP06} virial estimator. We preferred \iona{Mg}{ii}-based estimates to \iona{C}{iv}-based estimates 
when calculating the SMBH masses, because the \iona{C}{iv} virial estimator is more uncertain.
We calculated the bolometric luminosities as $L_{\rm Bol}=5.15\lambda L_{\lambda}(3000\textup{\AA})$ for \iona{Mg}{ii}-based
estimates or $L_{\rm Bol}=3.81\lambda L_{\lambda}(1350\textup{\AA})$ for \iona{C}{iv}-based estimates \citep{Shen2011}.
Of the seven \mbox{X-ray} weak WLQs in sample B+C, all but one (J1350$+$2618) have \iona{Mg}{ii}-based virial SMBH masses.

The $L_{\rm Bol}/L_{\rm Edd}$ values of our seven \mbox{X-ray} weak WLQs in sample B+C range from 0.06 to 0.79,
with a median value of 0.16. A Kolmogorov-Smirnov test indicates no significant difference between the $L_{\rm Bol}/L_{\rm Edd}$ values for
the seven \mbox{X-ray} weak WLQs and the 406 non-WLQs in sample B+C ($p=0.44$), which does not support the high $L_{\rm Bol}/L_{\rm Edd}$ scenario
for WLQs. However, we note that the \iona{Mg}{ii}- or \iona{C}{iv}-based virial masses are less reliable as
compared to H$\beta$-based virial masses, and the systematic uncertainties associated with the virial estimates might be
as large as $\ga0.4$~dex \citep[e.g.,][]{Shen2011}. Furthermore, virial estimates are most likely to fail at high Eddington ratios
\citep[e.g.,][]{Marconi2008,Marconi2009,NM2010}. Thus, the $L_{\rm Bol}/L_{\rm Edd}$ values derived from \iona{Mg}{ii}- or
\iona{C}{iv}-based estimates may be highly uncertain. Near-infrared spectroscopy covering the H$\beta$ line is needed to provide
more accurate $L_{\rm Bol}/L_{\rm Edd}$ estimates.

In sample A, there are 60 WLQs, 35 of which are \mbox{X-ray} detected and 25 of which are \mbox{X-ray}
undetected. Utilizing the Kaplan-Meier estimator in ASURV, we found the \mbox{X-ray} weak fraction within the 60 WLQs to be
\mbox{$35\pm8\%$}, which is consistent with that derived from sample B+C.
Of the nine detected \mbox{X-ray} weak WLQs in sample A, six are likely \mbox{X-ray} absorbed with
$\Gamma_{\rm eff}$ values ranging from $-0.8$ to 1.1, and two (both included in sample B+C) are potentially
\mbox{X-ray} unabsorbed. The other WLQ (included in sample B+C) is undetected in both the soft and hard bands, and thus
it is unclear if it is \mbox{X-ray} absorbed.

\subsubsection{\mbox{X-ray} Weak Red Quasars} \label{sec:red}

In addition to WLQs, there exists a population of red type 1 quasars that can be \mbox{X-ray} weak \citep[e.g.,][]{Wilkes2005,Hall2006}.
Previous \mbox{X-ray} studies have demonstrated that quasars with the reddest optical colors at their redshifts are more likely
to be \mbox{X-ray} absorbed than typical quasars \citep[e.g.,][]{Wilkes2005}, although some of the reddest quasars
may show no evidence of \mbox{X-ray} absorption or \mbox{X-ray} weakness \citep[e.g.,][]{Hall2006}. These optically red quasars may
be \mbox{X-ray} obscured by dusty gas, which also extincts the UV continuum, perhaps from a starburst disk surrounding their accreting
SMBHs \citep[e.g.,][]{Hickox2018}.

Broadband photometric studies of SDSS quasars have revealed a strong dependence of quasar optical/UV colors upon redshift
\citep[e.g.,][]{Richards2001}, and the relative $g-i$ color, $\Delta(g-i)$, 
is a useful redshift-independent indicator of the optical/UV spectral shape \citep[e.g.,][]{Richards2003}.
For a quasar at a given redshift, $\Delta(g-i)$ is defined as the difference between the $g-i$ color of that quasar
and the modal $g-i$ value of quasars at that redshift.
$\Delta(g-i)>0$ indicates a redder-than-average continuum, while $\Delta(g-i)<0$ indicates a bluer-than-average continuum.
The $\Delta(g-i)$ distribution for SDSS quasars is roughly Gaussian but has a distinct red tail with
$\Delta(g-i)\ga0.2$, which is indicative of dust reddening \citep{Richards2003}. Based on the $\Delta(g-i)$ distribution,
\citet{Hall2006} adopted $\Delta(g-i)>0.2$ to select red quasars for \mbox{X-ray} study.

The $\Delta(g-i)$ values of the 26 \mbox{X-ray} weak quasars in our sample B+C range from $-0.20$ to 0.82,
with a median value of 0.11 and a mean value of 0.17. Following \citet{Hall2006}, we adopted $\Delta(g-i)=0.2$
to be the threshold separating red and normal quasars in this paper.
Given this criterion, sample B+C contains eight \mbox{X-ray} weak red quasars, with $\Delta(g-i)$ values ranging
from 0.29 to 0.82. We note that although the $\Delta(g-i)>0.2$ criterion was designed to select dust-reddened
quasars, the possibility of selecting unreddened quasars (with intrinsically red continua) cannot be entirely excluded.
However, given that all our \mbox{X-ray} weak red quasars in sample B+C have \mbox{$\Delta(g-i)\ga0.3$}, they are likely to be dominated
by dust-reddened quasars \citep[e.g.,][]{Richards2003,Hall2006,Krawczyk2015}.

Given our red quasar definition of $\Delta(g-i)>0.2$, sample B+C contains 63 red quasars, eight of which are
\mbox{X-ray} weak, corresponding to an \mbox{X-ray} weak fraction of $13_{- 3}^{+ 5}\%$.
By contrast, only $ 5.0_{- 0.9}^{+ 1.4}\%$ (18/363) of non-red quasars are \mbox{X-ray} weak.
If we exclude red WLQs, the \mbox{X-ray} weak fraction for red and non-red quasars within sample B+C are
$ 9.3_{- 2.6}^{+ 5.5}\%$ (5/54) and $ 4.0_{- 0.8}^{+ 1.3}\%$ (14/352), respectively.
These results suggest that red quasars are likely to be another population of \mbox{X-ray} weak quasars in addition to WLQs,
although red quasars apparently have a smaller \mbox{X-ray} weak fraction than WLQs.

Of the eight \mbox{X-ray} weak red quasars, three \linebreak (J0039$+$0051, J1219$+$1244 and J1338$+$J2922)
are red WLQs with \mbox{\iona{C}{iv} REW $\rm {<12~\AA}$}. J0039$+$0051 and J1338$+$J2922 are likely \mbox{X-ray} absorbed with
$\Gamma_{\rm eff}=-0.8_{-3.3}^{+1.4}$ and $1.1_{-1.0}^{+0.9}$, respectively. Interestingly, J0039$+$0051 is the reddest
\mbox{($\Delta(g-i)=0.82$)} \mbox{X-ray} weak quasar in sample B+C, while J1338$+$J2922 has the highest Eddington ratio
($L_{\rm Bol}/L_{\rm Edd}=0.79$; see Section~\ref{sec:WLQ}) among \mbox{X-ray} weak WLQs in sample B+C. 
The other object, J1219$+$1244, has a fixed photon index of $\Gamma_{\rm eff}=1.8$ (see Section~\ref{sec:WLQ}).

Of the remaining five \mbox{X-ray} weak red quasars, four are likely \mbox{X-ray} absorbed, as their
$\Gamma_{\rm eff}$ values range from $-1.2$ to 1.1. The other quasar, J1522$+$0836, is undetected in both the soft and hard bands and
has a fixed photon index of $\Gamma_{\rm eff}=1.8$. Since dust causes optical/UV reddening and gas causes \mbox{X-ray} absorption,
the reddened optical/UV color together with absorbed \mbox{X-ray} emission indicate that these \mbox{X-ray} weak red quasars may
have optical/UV and \mbox{X-ray} absorption from dusty gas \citep{Hall2006}, perhaps from a starburst disk
\citep[e.g.,][]{Thompson2005,Ballantyne2008}.

We note that J1002$+$0203, the second reddest \mbox{($\Delta(g-i)=0.74$)}
\mbox{X-ray} weak quasar in sample B+C, shows relatively weak \iona{C}{iv} emission with \mbox{REW $\rm {=19~\AA}$}
(although not satisfying our WLQ criterion). It has 27 broad-band counts and is detected in all three \mbox{X-ray} bands.
J1002$+$0203 has a small photon index of $\Gamma_{\rm eff}=1.1$ and a high Eddington ratio of $L_{\rm Bol}/L_{\rm Edd}=0.59$.
Thus, the \mbox{X-ray} weakness of this quasar might also be explained via the thick inner disk scenario for WLQs
\citep[e.g.,][]{Luo2015,Ni2018}. 

Although their small $\Gamma_{\rm eff}$ values suggest \mbox{X-ray} absorption, these \mbox{X-ray} weak red quasars have
only \mbox{2--27} photon counts in the broad band, and thus the uncertainties on the effective photon indices $\Gamma_{\rm eff}$
are large. In addition, there are five \mbox{X-ray} weak red quasars (J0039$+$0051, J0228$+$0040, J1219$+$1244, J1239$+$1221,
and J1522$+$0836) that are mini-BAL quasars (see Section~\ref{sec:unc} below), and their \mbox{X-ray} weakness might also
be related to their mini-BAL features, although mini-BALs do not necessarily cause weak \mbox{X-ray} emission.

In sample A, there are 193 red quasars, 147 of which are \mbox{X-ray} detected and 46 of which are
\mbox{X-ray} undetected. Utilizing the Kaplan-Meier estimator in ASURV, we found the \mbox{X-ray} weak fraction within the 193 red
quasars to be \mbox{$12\pm3\%$}, which is consistent with that derived from sample B+C. 
Of the ten detected \mbox{X-ray} weak red non-WLQs in sample A, eight are likely \mbox{X-ray} absorbed and
their $\Gamma_{\rm eff}$ values range from $-1.2$ to 1.1. Another one (J1237$+$6203; not included in sample B+C)
is potentially \mbox{X-ray} unabsorbed with $\Gamma_{\rm eff}=1.6\pm0.4$. Its large $\Gamma_{\rm eff}$ value suggests
that this red non-WLQ is a candidate for an intrinsically \mbox{X-ray} weak quasar. The other one (J1522$+$0836;
included in sample B+C) is undetected in both the soft and hard bands, and thus it is unclear whether this object is \mbox{X-ray}
absorbed.

\subsubsection{Unclassified \mbox{X-ray} Weak Quasars} \label{sec:unc}

\begin{figure*}[htb!]
\centerline{
\includegraphics[scale=1.0]{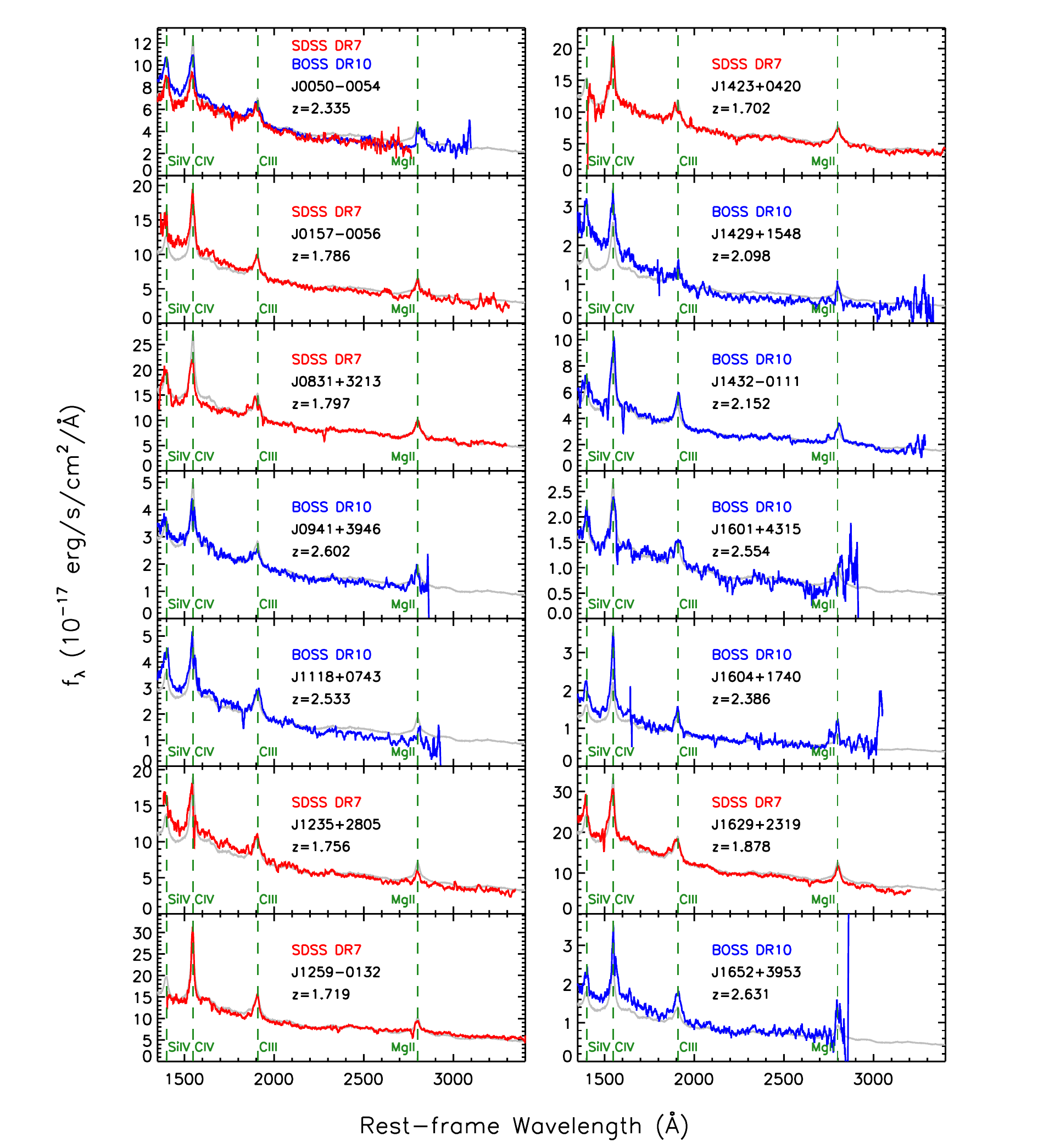}
}
\caption{SDSS optical/UV spectra (smoothed by a 20 pixel boxcar) for the 14 \mbox{X-ray} weak quasars in the unclassified
category (red curves for SDSS DR7 spectra and blue curves for BOSS DR10 spectra). The gray curve in each panel shows the SDSS
composite quasar spectrum \citep{Vanden Berk2001} for comparison. SDSS Data Release number, name, and redshift for each object
are listed.
\label{fig:sp_unc} }
\end{figure*}

In addition to the seven WLQs and five red non-WLQs, there are another 14 objects in sample B+C
that are \mbox{X-ray} weak. They do not have substantially unusual spectral features like WLQs or red quasars, and the reasons
for their \mbox{X-ray} weakness remain unclear. We consider an \mbox{X-ray} weak quasar that is neither a WLQ nor a red quasar
to be in the unclassified category (see Table~\ref{tbl:weak}). Figure~\ref{fig:sp_unc} shows the SDSS spectra of the 14
unclassified \mbox{X-ray} weak quasars in sample B+C. We proceed by enumerating some possible explanations for their weak
\mbox{X-ray} emission.

\begin{enumerate}
\item
One possible explanation for their \mbox{X-ray} weakness is relatively weak \iona{C}{iv} line emission. There are five
\mbox{X-ray} weak objects (J0050$-$0054, J0157$-$0056, J0941$+$3946, J1235$+$2805, and J1629$+$2319)
in the unclassified category that have \iona{C}{iv} REW
ranging from 16 to 25~\AA. Their relatively weak emission lines (although not satisfying our WLQ definition of \iona{C}{iv} REW
$\rm {<16~\AA}$) may explain their \mbox{X-ray} weakness following Section~\ref{sec:WLQ}. However, we note that a relatively weak
\iona{C}{iv} emission line does not necessarily lead to weak \mbox{X-ray} emission. For example, of the 46 quasars with
\iona{C}{iv} REW between 16 and 25~\AA\ in Sample B+C, 37 ($80_{- 7}^{+ 5}\%$) are \mbox{X-ray} normal (see also
section 5.2 of \citealt{Ni2018}). Therefore, it is just one possibility that the \mbox{X-ray} weakness of these five objects
may be related to their relatively weak \iona{C}{iv} emission lines. 
 
\item
The second explanation is relatively red optical/UV color following Section~\ref{sec:red}. One unclassified \mbox{X-ray} weak object,
J1259$-$0132, has $\Delta(g-i)=0.11$. Visual inspection of its SDSS spectrum (see Figure~\ref{fig:sp_unc}) confirmed that
J1259$-$0132 has a slightly redder-than-average continuum, which may explain its \mbox{X-ray} weakness. We note that like the relatively weak
\iona{C}{iv} emission line, a relatively red continuum does not necessarily cause weak \mbox{X-ray} emission either.
For example, of the 29 quasars with $0.1<\Delta(g-i)<0.2$ and visually confirmed redder-than-average continua in sample B+C,
27 ($93_{- 8}^{+ 2}\%$) are \mbox{X-ray} normal.

\item
Another possible explanation is mini-BALs, which are absorption features in between those of the traditional
BALs (trough width $\ge2,000$~\mbox{km~s$^{-1}$}) and narrow absorption lines (NALs; trough width $<500$~\mbox{km~s$^{-1}$}).
Mini-BAL quasars have $\Delta\alpha_{\rm OX}$ values intermediate between those of BAL and non-BAL quasars
\citep[e.g.,][]{Gibson2009b,Wu2010}, and some of them show \mbox{X-ray} absorption similar to that of BAL quasars \citep[e.g.,][]{Gallagher2002}.

Since we adopted the \citet{Trump2006} ${\rm AI}>0$ definition (see Section~\ref{sec:BAL}),
which requires a minimum trough width of 1,000~\mbox{km~s$^{-1}$}, to identify BAL features,
we defined mini-BALs using the following equation:   
\begin{eqnarray}
{\rm AI}_{\rm mini} &\equiv& \int^{29,000}_{0}(1-f(v))C'~dv.
\end{eqnarray}
Similar to the AI definition, in this equation $f(v)$ is the continuum-normalized flux density, and the value of $C'$ is
initially set to zero; it is set to 1 whenever the $f(v)$ has been continuously less than 0.9 for a trough width of
$500-1,000$~\mbox{km~s$^{-1}$}. \citet{Wu2010} suggested that mini-BALs with trough widths of $500-1,000$~\mbox{km~s$^{-1}$}
appear to be related to $\Delta\alpha_{\rm OX}$. We consider a sample object with ${\rm AI}_{\rm mini}>0$ a mini-BAL quasar. 
There are five unclassified \mbox{X-ray} weak quasars (J1423$+$0420, J1432$-$0111, J1601$+$4315,
J1629$+$2319,\footnote{J1629$+$2319 has a relatively weak \iona{C}{iv} emission line with REW $\rm {=24~\AA}$,
which may also explain its \mbox{X-ray} weakness.} and J1652$+$3953) in sample B+C
that are mini-BAL quasars, and their \mbox{X-ray} weakness may be explained by their mini-BAL features. However, we caution
that it is just one possibility that the \mbox{X-ray} weakness of these five objects may be related to their
mini-BAL features, as mini-BAL quasars do not appear to be an \mbox{X-ray} weak population in general. For example, among
the 156 mini-BAL quasars in sample B+C given our adopted definition, 144 ($92_{- 3}^{+ 2}\%$) are
\mbox{X-ray} normal. 

\item
The fourth possible explanation is NALs. Visual inspection of the SDSS spectra (see Figure~\ref{fig:sp_unc}) indicated that
J1118$+$0743 appears to exhibit narrow \iona{Al}{iii}~$\lambda1857$ absorption. Its \mbox{X-ray} weakness may be ascribed to
absorption, although like mini-BALs, NAL features do not necessarily lead to weak \mbox{X-ray} emission either
\citep[e.g.,][]{Ganguly2001}.

\item
The fifth explanation is \mbox{X-ray} variability, and we consider this the most plausible explanation
for the majority of the unclassified \mbox{X-ray} weak quasars. Such variable quasars may include quasars that vary strongly in the
\mbox{X-ray} band but not in the optical/UV, ``changing-look'' quasars, and quasars that show BAL disappearance or emergence.

Although most type 1 quasars show \mbox{X-ray} variability by factors of less than two \citep[e.g.,][]{Yang2016}, there is
a rare population of extremely \mbox{X-ray} variable quasars (e.g., PG 1211+143: \citealt{Bachev2009}; PG 0844+349:
\citealt{Gallo2011}; \citealt{GB2012}; PHL~1092: \citealt{Miniutti2012}; \citealt{Liu2019}; \citealt{Timlin2020})
which can vary by factors of larger than 10 in \mbox{X-ray}s but show little variation in the optical/UV bands.
They become remarkably \mbox{X-ray} weak in the low \mbox{X-ray} flux state,
and their \mbox{X-ray} weakness may be due to partial-covering absorption or inner-disk reflection
\citep[e.g.,][]{Bachev2009,Gallo2011,Miniutti2012,Liu2019,Ni2020}.
We note that such \mbox{X-ray} variability is not expected to have an \mbox{X-ray} strong state,
consistent with observations to date.

There also exist a few ``changing-look'' quasars that show strong long-term flux variability in both the optical/UV and
\mbox{X-ray} bands \citep[e.g.,][]{LaMassa2015}. If we calculate their $\alpha_{\rm OX}$ values based on non-simultaneous
\mbox{X-ray} and optical/UV observations, we may consider them to be \mbox{X-ray} weak quasars whenever their \mbox{X-ray}
dim states are combined with optically bright states.
Conversely, such quasars may become \mbox{X-ray} strong if their \mbox{X-ray}
bright states are combined with optically dim states.

There is another rare population of quasars showing emerging or disappearing
BALs \citep[e.g.][]{McGraw2017,De Cicco2018,Rogerson2018,Sameer2019}. These objects may be considered \mbox{X-ray} weak quasars
if they are non-BAL quasars when their optical/UV data are taken, but happen to exhibit BAL features, which cause weak
\mbox{X-ray} emission, when their \mbox{X-ray} data are taken.
The frequency of such highly variable \mbox{X-ray} absorption
in quasars with emerging or disappearing BALs is presently poorly constrained.
There is no \mbox{X-ray} strong state for this variability.

The SDSS and {\it Chandra} observation dates for the 14 sample B+C objects in the unclassified \mbox{X-ray} weak category
are listed in Table~7. It is clear that their SDSS optical/UV spectra and {\it Chandra} \mbox{X-ray} data are not simultaneous but
separated by $\approx1\textrm{--}4$ years in the rest frame. Thus, their \mbox{X-ray} weakness is possibly due to \mbox{X-ray}
and/or optical/UV spectral variability effects. Based on over 15,000 SDSS quasars at $z > 1.680$ matched between DR7 and
\mbox{DR9+DR10}, \citet{Rogerson2018} studied the emergence and disappearance of BAL features. They estimated the fraction of
non-BAL quasars turning into BAL quasars to be $0.59\%\pm0.12\%$ over timescales of \mbox{1--3} years in the rest frame. As the time 
separations between the \mbox{X-ray} and optical/UV observations for our sample B+C objects are comparable to the timescales of BAL 
variability studied in \citet{Rogerson2018}, we expect to see \mbox{2--3} sample B+C quasars exhibiting non-BAL-to-BAL transitions,
probably included in the 14 unclassified \mbox{X-ray} weak objects.

\begin{deluxetable}{lccc}
\tabletypesize{\scriptsize}
\tablecaption{SDSS and {\it Chandra} Observational Dates for Unclassified \mbox{X-ray} Weak Quasars in Sample B+C}
\tablehead{
\colhead{}  &
\colhead{}  &
\colhead{}  &
\colhead{Rest-frame} \\
\colhead{SDSS Name}  &
\colhead{SDSS Date}  &
\colhead{{\it Chandra} Date} &
\colhead{Separation (year)}
}
\startdata
$005018.84-005438.0$ & $10/17/2001$ & $09/02/2004$ & $ 0.8$ \\
$015704.11-005657.5$ & $10/17/2001$ & $11/27/2014$ & $ 4.4$ \\
$083116.62+321329.6$ & $12/12/2002$ & $12/22/2007$ & $ 1.8$ \\
$094138.06+394630.2$ & $03/09/2011$ & $01/16/2008$ & $ 1.1$ \\
$111828.31+074300.1$ & $01/03/2012$ & $01/31/2008$ & $ 1.4$ \\
$123559.06+280550.9$ & $12/25/2005$ & $01/14/2001$ & $ 1.8$ \\
$125923.50-013234.9$ & $05/30/2000$ & $02/28/2009$ & $ 3.1$ \\
$142339.87+042041.1$ & $05/20/2001$ & $12/15/2012$ & $ 3.8$ \\
$142921.75+154841.4$ & $04/18/2012$ & $12/06/2013$ & $ 0.5$ \\
$143212.69-011109.7$ & $02/26/2011$ & $03/31/2000$ & $ 3.2$ \\
$160154.45+431519.6$ & $06/18/2012$ & $10/07/2003$ & $ 3.2$ \\
$160408.26+174042.4$ & $05/22/2010$ & $07/25/2004$ & $ 2.1$ \\
$162922.87+231958.2$ & $08/09/2004$ & $10/20/2013$ & $ 3.0$ \\
$165209.44+395348.8$ & $05/27/2012$ & $05/19/2014$ & $ 0.7$ \\
\hline
\enddata
\label{tbl:date}
\end{deluxetable}

\end{enumerate}

Given our definition of potentially \mbox{X-ray} absorbed quasars ($\Gamma_{\rm eff}<1.26$), nine of these 14
unclassified \mbox{X-ray} weak quasars are likely \mbox{X-ray} absorbed, with their $\Gamma_{\rm eff}$ values ranging from $-0.4$
to 1.0. Four quasars (J1118$+$0743, J1429$+$1548, J1601$+$4315, and J1629$+$2319) are potentially \mbox{X-ray} unabsorbed
with $\Gamma_{\rm eff}=1.3\pm0.50$, $1.4_{-0.6}^{+0.7}$, $2.1_{-1.3}^{+1.0}$,
and $2.3_{-1.4}^{+0.9}$ respectively; these are candidates for intrinsically \mbox{X-ray} weak quasars.
The remaining quasar, J0941$+$3946, is undetected in both the soft and hard bands, and thus it is unclear whether this object
is \mbox{X-ray} absorbed. We caution that these 14 unclassified \mbox{X-ray} weak quasars have only 2--19 photon counts
in the broad band, leading to substantial uncertainties on their $\Gamma_{\rm eff}$ values.

In sample A, there are 45 unclassified \mbox{X-ray} weak quasars, nine of which have relatively weak emission
lines with \iona{C}{iv} REW ranging from 16 to 25~\AA, two of which have $0.1<\Delta(g-i)<0.2$ and visually confirmed
redder-than-average continua, and 12 of which are mini-BAL quasars. All these 45 unclassified
\mbox{X-ray} weak quasars are \mbox{X-ray} detected. Among these quasars, 37 are likely \mbox{X-ray} absorbed with $\Gamma_{\rm eff}$
values ranging from $-1.4$ to 1.2, and seven (three not included in sample B+C) are potentially
\mbox{X-ray} unabsorbed, with their $\Gamma_{\rm eff}$ values ranging from 1.3 to 2.3.
These seven objects are candidates for intrinsically \mbox{X-ray}
weak quasars. The other quasar (included in sample B+C) is undetected in both the soft and hard bands,
and thus it is unclear if this object is \mbox{X-ray} absorbed.

\section{Summary and Future Work} \label{sec:SF}
We have investigated systematically the \mbox{X-ray} properties of a large sample of SDSS quasars.
We constrained the fraction of \mbox{X-ray} weak quasars, and discussed the possible causes of quasar \mbox{X-ray} weakness.
The main points from this work are the following:

\begin{enumerate}
\item
After removal of BAL and RL quasars, we selected the final sample, or sample A, of 1825 SDSS DR7 and DR10 quasars
with {\it Chandra} archival \mbox{X-ray} observations (Section~\ref{sec:SA}). We measured their \mbox{X-ray} properties
(Section~\ref{sec:X}), calculated the $\alpha_{\rm OX}$ parameter, and investigated the
\mbox{$\alpha_{\rm OX}$--$L_{\rm 2500~{\textup{\AA}}}$} relation (Section~\ref{sec:al}).

\item
The $\Delta\alpha_{\rm OX}$ distribution for sample A has a clear negative tail, suggesting the existence of a population of
\mbox{X-ray} weak quasars. The fraction of \mbox{X-ray} weak quasars ($\Delta\alpha_{\rm OX}\le-0.3$) is $5.8\pm0.7\%$, and
quasars that are \mbox{X-ray} weak by factors of 10 and 20 represent \mbox{$2.7\pm0.5\%$} and \mbox{$1.3\pm0.3\%$}
of the population, respectively. See Section~\ref{sec:frac}.

\item
We constructed two subsamples (samples B and C), without direct reference to their \mbox{X-ray} properties, that contain
218 DR7 and 208 DR10 quasars, respectively. Both subsamples are composed of \mbox{X-ray} detections only, and the fractions of \mbox{X-ray} weak
quasars among these two subsamples are consistent with that in sample A. See Section~\ref{sec:frac_BC}.

\item
We note that the fraction of \mbox{X-ray} weak quasars likely has a luminosity dependence. The fractions derived in this study are
mostly applicable to quasar samples sharing similar
properties to our SDSS quasars here (Table~1), and extra caution is needed if quoting the fractions for quasars with
substantially different luminosities. See Section~\ref{sec:fxw}.

\item
WLQs (\iona{C}{iv} REW $<16~{\textup{\AA}}$) represent one population of \mbox{X-ray} weak quasars, and their \mbox{X-ray} weak fraction
($35_{- 9}^{+12}\%$) is significantly higher than that of non-WLQs. See Section~\ref{sec:WLQ}.

\item
Red quasars ($\Delta(g-i)>0.2$) are likely to be another population of \mbox{X-ray} weak quasars, and their \mbox{X-ray} weak fraction 
($13_{- 3}^{+ 5}\%$) is considerably higher than that of normal quasars. See Section~\ref{sec:red}.

\item
We provide several possible explanations for the \mbox{X-ray} weakness of quasars in the unclassified category: relatively
weak \iona{C}{iv} emission lines, relatively red optical/UV continua, mini-BALs, NALs, and \mbox{X-ray} variability.
We consider \mbox{X-ray} variability the most plausible explanation; such quasars include quasars that vary strongly in the
\mbox{X-ray} band but not in the optical/UV, ``changing-look'' quasars, and quasars that show BAL disappearance or emergence.
See Section~\ref{sec:unc}.

\end{enumerate}

Further work is needed to improve the results of this study. Deeper radio observations, e.g., with the Very Large Array Sky Survey
\citep[VLASS, e.g.,][]{Lacy2019,Lacy2020} will be helpful to discriminate $R \ge 10$ vs. $R<10$ for the optically faint quasars
(especially the post--DR7 quasars), thus removing non-RQ objects with possibly enhanced \mbox{X-ray} emission. Deeper \mbox{X-ray}
observations, e.g., {\it Chandra} observations with typical individual exposure times of \mbox{$\approx30$~ks} (estimated from
the current median exposure time and source counts) are required to better constrain the spectral shapes of the detected sources, 
thus helping distinguish more reliably between absorbed and unabsorbed quasars. 

Additionally, although \mbox{X-ray} weak quasars constitute only a small fraction of the non-BAL quasar population, they may provide us
with new insights into the SMBH disk-corona system and BELR. Deeper targeted \mbox{X-ray} observations of our \mbox{X-ray} weak quasars,
especially those candidates for intrinsically \mbox{X-ray} weak quasars and unclassified \mbox{X-ray} weak quasars, are necessary to assess
their physical nature. For example, {\it Chandra} observations with exposure times of \mbox{$\approx100$~ks} will on average yield
\mbox{$\approx50$} broad-band counts for the \mbox{X-ray} weak quasars in sample B+C.

Our study here will benefit from a larger sample of \mbox{X-ray} quasars, to increase the
statistical significance of the results and to identify more \mbox{X-ray} weak quasars.
The SDSS Data Release 16 quasar catalog \citep[DR16Q;][]{Lyke2020}
contains $750,426$ spectroscopically confirmed quasars. 
Combining the {\it Chandra} archive with the DR16Q will extend the \mbox{X-ray} quasar study to an even larger sample.
Furthermore, the {\it XMM-Newton} serendipitous source catalogs \citep[e.g.,][]{Watson2009,Rosen2016} are good complements to the
{\it Chandra} archival \mbox{X-ray} data. The latest 4XMM-DR9 catalog covers a sky area of 1152 deg$^{2}$, and contains $810,795$ detections
comprising $550,124$ unique \mbox{X-ray} sources.\footnote{\url{http://xmmssc.irap.omp.eu/Catalogue/4XMM-DR9/4XMM_DR9.html}} With an
estimated $20,000\textrm{--}25,000$ matches in the DR16Q catalog, the 4XMM-DR9 catalog can significantly increase the number of \mbox{X-ray} quasars.

Finally, {\it eROSITA} (extended ROentgen Survey with an Imaging Telescope Array; \citealt{Merloni2012}), 
a Russian-German space mission, will discover many more \mbox{X-ray} AGNs and quasars in the near future.\footnote{\url{http:
//www.mpe.mpg.de/eROSITA}} Launched on 13 July 2019, {\it eROSITA} is performing a medium-depth \mbox{X-ray} all-sky survey over the next four years.
{\it eROSITA} is to map the entire sky in the soft \mbox{X-ray} band \mbox{(0.5--2~keV)} at a level $>20$ times more sensitive than the {\it ROSAT}
all sky survey, as well as in the hard \mbox{X-ray} band \mbox{(2--10~keV)} which provides the first imaging all sky survey at these higher
\mbox{X-ray} energies. The 4-year {\it eROSITA} all-sky survey is expected to detect $\approx3$ million AGNs and quasars. However, the expected 
sensitivity of the 4-year all-sky survey is \mbox{$\approx1\times10^{-14}$~\flux} in the soft band (\mbox{0.5--2~keV}), which equals the median 
\mbox{0.5--2~keV} flux of sample B+C quasars. The \mbox{0.5--2~keV} fluxes of the 26 \mbox{X-ray} weak quasars in sample B+C
range from \mbox{$\approx1\times10^{-16}$~\flux} to \mbox{$4\times10^{-15}$~\flux}, with median and mean values of \mbox{$\approx2\times10^{-15}$~\flux}. 
Thus, the {\it eROSITA} \mbox{all-sky} survey will probably be unable to enlarge significantly the sample size or
unravel the nature of these exceptionally \mbox{X-ray} weak quasars identified here.

\acknowledgments
We thank the referee for the helpful comments.
XP, BL, and HL acknowledge financial support from the National Natural Science Foundation of China grants 11673010 and 11991053,
National Key R\&D Program of China grant 2016YFA0400702, the Natural Science Foundation of Jiangsu Province (Grant No. BK20150870),
and Jiangsu Planned Projects for Postdoctoral Research Funds grant 1601060B.
WNB and JDT acknowledge support from NASA ADP grant 80NSSC18K0878,
Chandra \mbox{X-ray} Center grant G08-19076X, and the V.M. Willaman Endowment.

\appendix

\section{J1457$+$2218: a Candidate for an Intrinsically \mbox{X-ray} Weak Quasar} \label{sec:Xspec}

One of the WLQs, J1457$+$2218, is a good candidate for an intrinsically \mbox{X-ray} weak quasar.
It has 70 broad-band counts in the source-extraction region: 47 in the soft band and 23 in the hard band, sufficient
for basic spectral fitting. Using the CIAO {\sc specextract} script, we extracted the broad-band spectrum of J1457$+$2218 from
a circular region with a radius of 3.3 pixels centered at the \mbox{X-ray} position, corresponding to a 90\% encircled-energy fraction.
The background spectrum was extracted from an annular region with an inner radius of 18.3 pixels and an outer radius of 53.3 pixels,
which corresponds to the source radius plus 15 and 50 pixels, respectively. Note that J1457$+$2218 is close to a galaxy cluster,
MS~1455.0+2232, which is the target of the {\it Chandra} observation, and the distance between their centers is 2.0\arcmin.
Thus, the source and background regions of J1457$+$2218 are likely to be contaminated by the diffuse source.
However, since the photon counts in the background region are relatively uniformly distributed and the spectral counts contain only
$\approx1$ background count, the nearby galaxy cluster does not affect our spectral analysis significantly.

We performed spectral fitting using XSPEC v12.9.0 \citep{Arnaud1996}. The $C$-statistic ({\tt cstat}) was used given the 
limited source counts \citep{Cash1979}. We first fit the spectrum with a power-law model modified by only the Galactic absorption
({\tt zpow*wabs}). The photon index of $\Gamma=1.6_{-0.3}^{+0.4}$ derived from this spectral fitting is
in agreement with $\Gamma_{\rm eff}=1.6_{-0.2}^{+0.3}$ (Table~\ref{tbl:WLQ}) derived from the band ratio.

\begin{figure*}[t]
\centerline{
\includegraphics[scale=0.5]{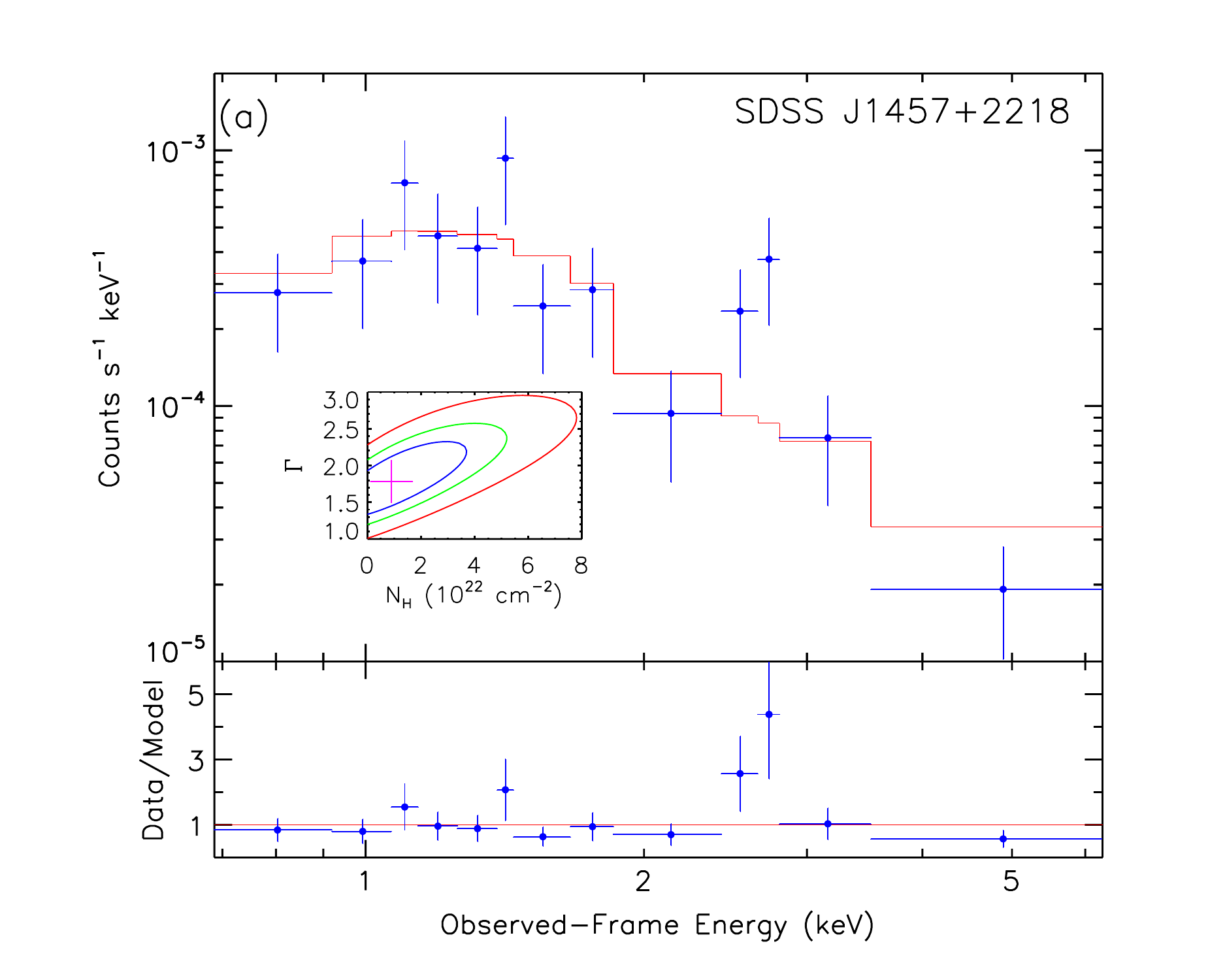}
\includegraphics[scale=0.5]{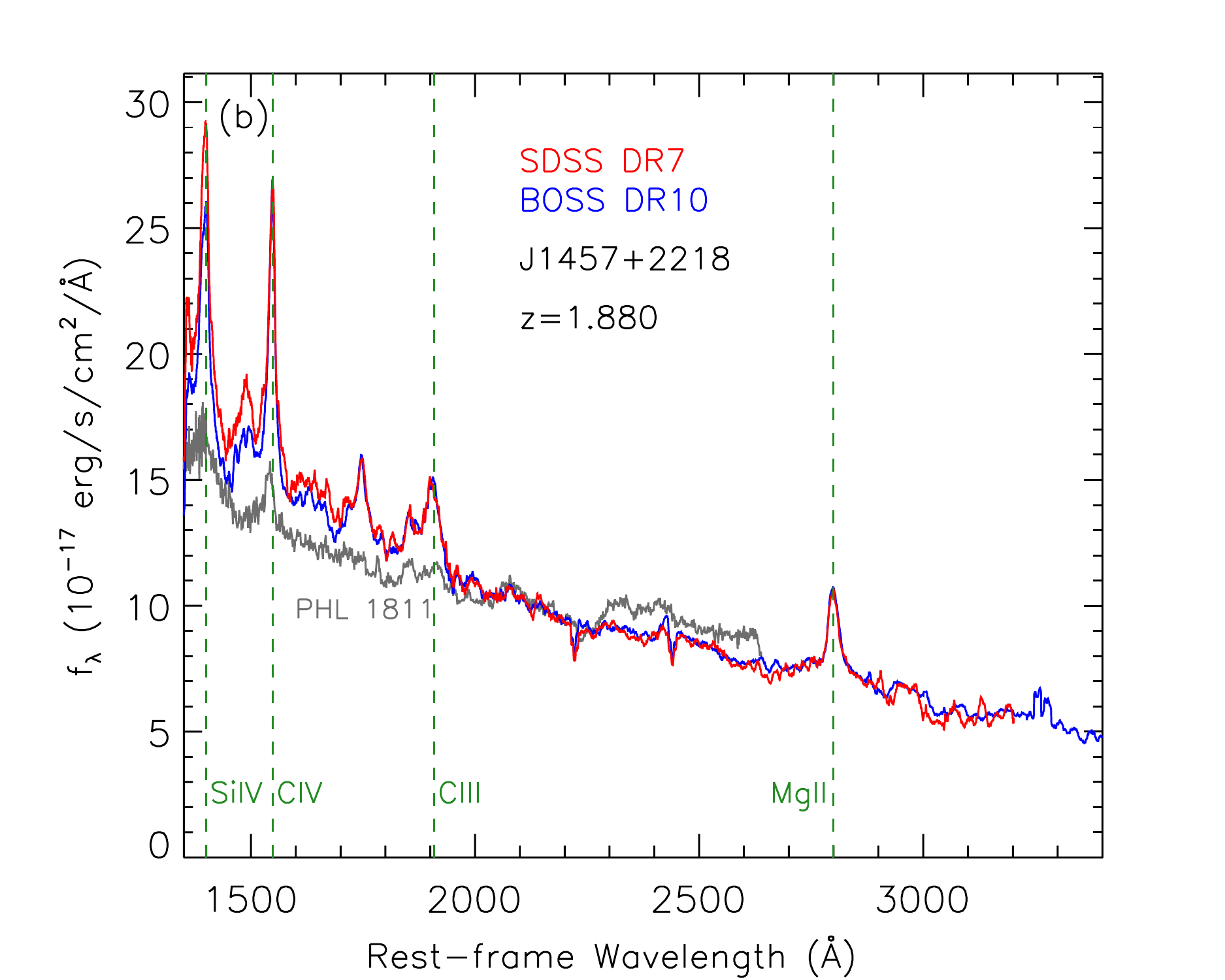}
}
\caption{(a) {\it Chandra} broad-band (0.5--7~keV) spectrum of J1457$+$2218 (blue) and the best-fit model with both Galactic and
intrinsic absorption (red). The spectrum was binned to a minimum of five counts per bin for display purposes. The inset shows
the best-fit values of $\Gamma$ and intrinsic $N_{\rm H}$ (magenta), and contours of $\Gamma$ vs. $N_{\rm H}$ at confidence
levels of 68\% (blue), 90\% (green), and 99\% (red), respectively. The bottom panel is the data divided by the model. 
(b) SDSS DR7 spectrum (red) and BOSS DR10 spectrum (blue) of J1457$+$2218. The gray curve shows the PHL~1811 spectrum
for comparison.
\label{fig:Xspec} }
\end{figure*}

We also fit the spectrum with a power-law model modified by both Galactic absorption and intrinsic absorption at the quasar
redshift ({\tt zpow*wabs*zwabs}). The best-fit values of the photon index and intrinsic absorption column density are
$\Gamma=1.8_{-0.3}^{+0.4}$ and $N_{\rm H}=0.90_{-0.89}^{+1.74}$ $\times10^{22}$~{cm$^{-2}$}, respectively.
The errors are quoted at a 68\% (1$\sigma$) confidence level. Figure~\ref{fig:Xspec}(a) shows the \mbox{X-ray} spectrum of J1457$+$2218
and the best-fit model. 
We note that there appears to be excess residuals at rest-frame $\approx7.6\textrm{--}8.1$~keV.
While the current {\it Chandra} effective exposure time is 88.4~ks, still deeper \mbox{X-ray} data are required to determine if
there is any line emission, which may be due to blueshifted iron emission.
It appears that J1457$+$2218 does not suffer from strong \mbox{X-ray} absorption.
J1457$+$2218 has $\Delta\alpha_{\rm OX}=-0.30$, which corresponds to an \mbox{X-ray} weakness factor of $f_{\rm weak}=6.1$,
and it remains \mbox{X-ray} weak by a factor of 4.7 after being corrected for the intrinsic absorption.
J1457$+$2218 is also RI with $R=42$ and may have some jet-linked \mbox{X-ray} emission, which would indicate even weaker
intrinsic coronal \mbox{X-ray} emission. The spectral fitting results suggest that J1457$+$2218 is probably an intrinsically
\mbox{X-ray} weak WLQ like PHL~1811. Figure~\ref{fig:Xspec}(b) shows the optical/UV spectra of J1457$+$2218 and PHL~1811
for comparison. J1457+2218 has overall stronger line emission than PHL~1811; for example, the SDSS (BOSS) {\rm \iona{C}{iv}} REW is
$\rm {22.0~{\textup{\AA}}}$ ($\rm {15.5~{\textup{\AA}}}$), while it is $\rm {4.7~{\textup{\AA}}}$ for PHL~1811.

\end{document}